\newcommand{\be}{\begin{equation}}
\newcommand{\bea}{\begin{eqnarray}}
\newcommand{\ba}{\begin{array}}
\newcommand{\ee}{\end{equation}}
\newcommand{\eea}{\end{eqnarray}}
\newcommand{\ea}{\end{array}}
\newcommand{\qa}{\alpha}
\newcommand{\qb}{\beta}
\newcommand{\qg}{\gamma}
\newcommand{\qG}{\Gamma}
\newcommand{\qd}{\delta}

\newcommand{\qe}{\varepsilon}

\newcommand{\qh}{\eta}

\newcommand{\qk}{\kappa}
\newcommand{\ql}{\lambda}
\newcommand{\qL}{\Lambda}

\newcommand{\qr}{\rho}
\newcommand{\qs}{\sigma}

\newcommand{\qt}{\tau}
\newcommand{\qf}{\varphi}
\newcommand{\qF}{\Phi}

\newcommand{\qj}{\psi}
\newcommand{\qJ}{\Psi}
\newcommand{\qo}{\omega}

\newcommand{\sgn}{{\rm sgn}}

\newcommand{\tr}{{\rm tr}\;}
\newcommand{\Tr}{{\rm Tr}\;}

\newcommand{\inv}{^{-1}}

\newcommand{\dagg}{^{\dag}}
\newcommand{\prt}{\partial}

\newcommand{\intd}[1]{\int \!\! d{#1} \;}

\newcommand{\intdxx}{\int \!\! d^{2}x \;}

\newcommand{\intit}{\int_{0}^{\beta}\!\! d\tau}
\newcommand{\pathint}[1]{\int\!\! {\cal D} #1 \;}

\newcommand{\fr}[2]{{\textstyle \frac{#1}{#2}}}
\newcommand{\half}{\mbox{$\textstyle \frac{1}{2}$}}

\newcommand{\naar}{\rightarrow}

\newcommand{\nn}{\nonumber}

\newcommand{\scez}{\setcounter{equation}{0}}

\newcommand{\ns}{\scez \section}
\renewcommand{\theequation}{\thesection .\arabic{equation}}

\newcommand{\pileft}{\stackrel{\scriptscriptstyle{\leftarrow}}{\pi}}
\newcommand{\nablaleft}{\stackrel{\scriptscriptstyle{\leftarrow}}{\nabla}}

\newcommand{\Ian}{{\rm I}^\alpha_n}
\newcommand{\Iamn}{{\rm I}^\alpha_{-n}}

\newcommand{\eff}{_{\rm eff}}
\newcommand{\tQ}{\widetilde{Q}}
\newcommand{\tI}{\tilde{I}}
\newcommand{\curl}{\nabla\!\times\!}

\documentstyle[preprint,aps,eqsecnum,epsf,prb, tighten]{revtex}
\begin{document}

\draft

\title{(Mis-)handling gauge invariance in the theory \\of the quantum Hall
 effect
I: \\ Unifying action and the $\nu=\half$ state}

\author{A.M.M. Pruisken, M.A. Baranov\cite{Misha}, B. \v{S}kori\'{c} }

\address{Institute for Theoretical Physics, University of Amsterdam,
Valckenierstraat 65, \\ 1018 XE Amsterdam, The Netherlands}

\date{October 9, 1998}
\maketitle

\begin{abstract}
\noindent
We propose a unifying theory for both the integral and fractional
quantum Hall regimes. This theory reconciles the Finkelstein approach
to localization and interaction effects with the topological issues of
an instanton vacuum and Chern-Simons gauge theory. We elaborate on the
microscopic origins of the effective action and unravel a new symmetry
in the problem with Coulomb interactions which we name ${\cal
F}$-invariance. This symmetry has a broad range of physical
consequences which will be the main topic of future analyses. In the
second half of this paper we compute the response of the theory to
electromagnetic perturbations at a tree level approximation. This is
applicable to the theory of ordinary metals as well as the composite
fermion approach to the half integer effect. Fluctuations in the
Chern-Simons gauge fields are found to be well behaved only when the theory
is ${\cal F}$-invariant.
\end{abstract}


\ns{Introduction}
It is well known that the quantum Hall effect exists only due to the
presence of random impurities \cite{c3}. Although one usually prefers to think
in terms of the pure incompressible states alone, the random
impurities problem in all its generality opens up a Pandora's box of
concepts and complex analyses. The integral quantum Hall regime is the
simplest and most widely studied example. The advances made in this
field have in fact very little to do with the Landau quantization of
pure states that one originally starts out from \cite{c3}. Actually, what is
really needed is topological ideas in quantum field theory (instanton
vacuum \cite{a2,a4}) in order to establish a unifying renormalization
theory for the many unrelated experimental phenomena that are observed in the
laboratory \cite{a8}. Some examples of these phenomena are weak localization in
weak magnetic fields and higher Landau levels \cite{a6}, 
the quantization of the
Hall conductance in strong magnetic fields \cite{c3}, the (critical) plateau
transitions which occur usually in the Landau band centers
\cite{a9,a11}, 
the problem of spin unresolved Landau levels etc. The
problem that needs to be addressed is characterized by a rich variety
of different cross-over length scales, whereas detailed experiments are often
difficult to interpret due to the limited range available in
temperature and sample-specific properties (long range versus short
range disorder \cite{a8} etc.).

To date, the Pandora's box of the integral quantum Hall regime is the
only one that has been opened. For the fractional quantum Hall regime,
which is believed to be a strongly correlated phenomenon, the impurity
problem has not yet even been formulated! Our theoretical
understanding has not progressed beyond that of the popularly studied
incompressible pure states as initiated by Laughlin \cite{c5}.

Nevertheless, there has been a long standing quest for a unifying
theory which would combine the basic effects of disorder and strong
correlation into a single renormalization group flow diagram of the
conductances \cite{b18}. The original attempts made in this direction
were solely motivated by the experiments which seemed to indicate that the
integral and fractional effects have very similar common features.

In this paper, and others that follow, we will lay the foundation and
construct the much sought after unifying theory. However, formulation
of this unifying theory heavily relies on advances made in the recent
literature on the fractional and integer quantum Hall effect. In
particular, we refer to the analysis of localization and interaction
made by two of the authors \cite{PruiskenBaranov}, in which
Finkelstein's effective sigma model theory \cite{b8} was
extended by including topological effects (instanton vacuum)
\cite{PruiskenBaranov}. 
In this
work it was shown that the interacting electron gas shares many of the
basic features which were previously found for free electrons, namely
asymptotic freedom in two dimensions and non-perturbative
renormalization by instantons. These results put the topological
concept of an instanton vacuum in an entirely different perspective of
many body theory, the consequences of which have yet remained largely
unexploited. Secondly, there is the Chern-Simons gauge field approach
\cite{Wilczek82,c11,c12,c13,c14,b5,c15}
which implicitly carries out Jain's composite fermion ideas
\cite{c10} and maps the fractional quantum Hall effect onto the
integral effect. We are specifically interested in the fermionic
Chern-Simons theory \cite{c12}, since the basic starting formulation (the
fermionic path integral) is the only one suitable for analyzing
disorder effects and, in particular, the above mentioned instanton
vacuum concept.

The effective action proposed in this paper essentially extends the
Finkelstein theory \cite{b8} to include the topological effects of an instanton
vacuum as well as Chern-Simons gauge theory. As one of the principal
features of our theory we shall show that most of the presently
accumulated knowledge on the quantum Hall regime can be derived from
our effective action by considering the extreme limits of {\em weak}
and {\em strong} coupling only. More specifically, the theory in weak
coupling describes the composite fermion approach to the half integral
effect. This will be the main subject in the second half of this
paper. The theory in strong coupling on the other hand describes the
Jain series for fractional quantization of the Hall conductance \cite{b30} and
also provides a microscopic theory of disordered chiral edge states
which generalizes and extends the previously introduced Luttinger
liquid description for edge states without disorder \cite{b27}.
Subsequent papers will report the strong coupling effects \cite{b31}.

It is interesting to note that the physics of both weak and strong
coupling is essentially a perturbative phenomenon from the instanton
vacuum point of view. Nevertheless, our results clearly show that our
effective action can be used to establish a much more ambitious theory
for the quantum Hall effect. It is possible to address and investigate
the consequences of the renormalizability (both perturbative and
non-perturbative) of the theory \cite{b32}. Further, this will provide the
necessary information on the global phase structure of the quantum
transport problem in the presence of random impurities.

In this paper we mainly explain the microscopic origins of the
effective action that we propose as the unifying theory for both
integral and fractional quantum Hall regimes. The analysis presented
is largely based on the insights we have accumulated by extensively
studying the free electron renormalization theory of the integral
effect. We will therefore refer to this analysis \cite{a4,c3} throughout
the course of this work. A second important reference which has been
critical in motivating this work is the above mentioned
renormalization group analysis of localization and interaction effects
in the quantum Hall regime \cite{PruiskenBaranov}. During the course of this analysis we
became aware of the incomplete nature of Finkelstein's pioneering work
on the subject. One of the major complications in Finkelstein's
approach is the $U(1)$ electrodynamic gauge invariance of the
theory. For example, no transparent and consistent way exists for
introducing external vector and scalar potentials and/or Chern-Simons
gauge fields. Subsequently this prevents one from using this theory as
a microscopic approach for the fractional Hall effect.

In order to construct a unifying theory we start out (Section~\ref{secQfield})
by considering the fermionic path integral in Matsubara frequency
representation. We then analyze in a step by step manner both the low
energy excitations in the problem ($Q$-fields) and the $U(1)$ gauge
invariance. Upon performing this exercise, we find that the $U(1)$
generators for the gauge fields and those for the $Q$-fields cover
distinctly different sectors in Matsubara frequency space, which
naively appear to be completely disconnected. Assuming this to be
true, one would not even consider using the effective action
approach. However, these two apparently distinct aspects are
integrally related to one another via Ward identities (obtained from
local $U(N)\!\times\!U(N)$ symmetry) \cite{b32}.
In order to establish this relationship, several concepts such as
`smallness' of the $Q$-fields and `${\cal F}$-algebra' are
introduced. These concepts are absolutely necessary for handling the
$U(1)$ gauge invariance of the problem. Additionally, they also
elucidate a hidden symmetry in the Finkelstein action which has
previously gone unnoticed. The symmetry we unravel has far reaching
consequences and plays a critical role in the development of the
unifying theory. We term this symmetry as ${\cal F}$-invariance since
we will be frequently referring to it in the rest of this work.

One of the important consequences for ordinary metals is that the
nature of quantum transport fundamentally changes depending on the
length scale being considered. At distances short relative to the
Debije screening length, transport is free particle like and
conservation laws are governed by Einstein's relation between
conduction and diffusion. At large distances, however, the metal is no
longer diffusive and the internally generated electric field due to
the Coulomb interactions enters into the transport equations. Our
theory provides these results by considering the gauge invariant
response at the tree level (Section~\ref{secexternal}).

Finally, in Section~\ref{secCS} we include the Chern-Simons 
statistical gauge fields in the action. As a first step towards
describing the fractional Hall regime, we consider the $\nu=1/2$
state. In this case it is sufficient to work with the statistical
gauge fields and external fields in the tree level
approximation. Additionally, we compute the contribution of the
Chern-Simons gauge fields to the specific heat. For the problem with
Coulomb interactions we find that the singularity structure of the
theory is not modified. This then demonstrates that the composite
fermion approach to the half integral effect is free of infrared
trouble. On the other hand, for a system with finite range
electron-electron interactions complications do arise. These aspects
are further discussed in Section~\ref{secinternal}.

We end this paper with a conclusion (Section~\ref{secconclusion}).

\ns{$Q$-field formalism; the fermionic path integral}
\label{secQfield}

\subsection{Introduction}
\label{secintro}

We are interested in the disorder average of the logarithm of the grand
canonical partition function $Z$,
\be
        Z=\tr e^{\qb(\mu N-H)}
\ee
with $\qb$ the inverse thermal energy, $\qb=(k_{\rm B} T)\inv$, $\mu$
the chemical potential, $N$
the number of electrons and $H$ the total energy of the system.
We consider a system of
two-dimensional electrons in a random potential $V(\vec x)$ and a static
magnetic field $B$ pointing along the positive $z$-axis.
We work in units where all lengths are expressed in terms of the magnetic
length $\ell\! =\!\sqrt{\fr{2\hbar}{eB}}$ and where 
$\hbar\! =\! 1$, $e\! =\! 1$. In these
units, the electron mass $m$ has the dimension of an inverse energy, while the
static magnetic field and the vector potential
are dimensionless, 
\bea
	m=m_{\rm SI}\cdot\ell^2/\hbar^2 \hskip0.5cm&;& \hskip0.5cm
	\vec A=\vec A_{\rm SI}\cdot e\ell/\hbar.
\eea
We write the vector potential as $\vec A^{\rm st}\! +\!\vec A$, where the
static part satisfies $\curl \vec A^{\rm st}=B\vec e_z$ and 
$\vec A$  represents the quantum fluctuations. In the units chosen
above, the magnetic field is normalized to $B\! =\! 2$. The fluctuations in
the scalar potential are denoted by $A_0$.
In the path integral formulation, the partition function 
for our system is written in the
following way
\bea
        Z &=& \int\!{\cal D}[\bar{\qj}\qj,A_\mu]\; 
	e^{S[\bar{\qj},\qj,A_\mu]} \\
	S[\bar{\qj},\qj,A_\mu] & = &
        \intit\intdxx \bar{\qj}(\vec x,\qt)[-\prt_{\qt}+iA_{\qt}(\vec x,\qt)
	+\mu-{\cal H}(\vec x)
        -V(\vec x)]\qj(\vec x,\qt) \nn\\
	& & -\half\intit\intd{^2 x d^2 x'}\bar{\qj}(\vec x,\qt)\qj(\vec x,\qt)
	U_0(\vec x,\vec x')\bar{\qj}(\vec x',\qt)\qj(\vec x',\qt).
\label{basicaction}
\eea
Here, the $\qj$ and $\bar{\qj}$ are Grassmann variables defined on the
imaginary time interval 
$\qt\in[0,\qb]$, with the fermionic antiperiodicity condition
$\qj(\vec x,\qb)\! =\! -\qj(\vec x,0)$. The $A_\mu$ are ordinary integration
variables with the bosonic boundary condition 
$A_\mu(\vec x,\qb)\! =\! A_\mu(\vec x,0)$.
The $U_0(\vec x,\vec x')$ is the Coulomb interaction and
${\cal H}$ is a differential operator acting to the left and to the right
\be
	{\cal H}=\fr{1}{2m} \pileft\cdot\vec\pi,
\ee
where $\pi$ is the covariant derivative,
\bea
	\vec\pi= -i\vec\nabla-\vec A^{\rm st}-\vec A & ; &
	\pileft= i\nablaleft-\vec A^{\rm st}-\vec A.
\eea
The Coulomb term is quartic in the fields $\qj$. We get rid of this
quartic form by performing a Hubbard-Stratonovich transformation,
introducing an extra path integration over a bosonic field 
$\ql(\vec x,\qt)$, the `plasmon field',
\bea
	&&\exp -\half\intd{\qt d^2 x d^2 x'}
	\bar{\qj}\qj(\vec x)U_0(\vec x,\vec x')\bar{\qj}\qj(\vec x') 
	\propto  \\
	&&\pathint{[\ql]}\exp\left[-\half\intd{\qt d^2 x d^2 x'}\ql(\vec x)
	U_0\inv(\vec x,\vec x')\ql(\vec x')
	+i\intd{\qt d^2 x}\ql\bar{\qj}\qj
	\right]. \nn
\eea
Here $U_0\inv$ stands for the matrix inverse of $U_0$.
In order to find the disorder average $\overline{\ln Z}$ we use the well
known {\em replica trick} \cite{a4,c3}. 
In the path integral formalism this amounts to the introduction of
replicated fields $\bar{\psi}^{\qa},\qj^{\qa}, \ql^{\qa}, A_\mu^\qa$ 
with $\qa\! =\! 1,\cdots,N_r$. The quantities $\mu$, $\qr$, $V$
and $A_{\mu}^{\rm st}$ are identical in all replicas. The replicated
partition function is given by
\bea
	Z &=& \int\!\prod_{\qg=1}^{N_r}{\cal D}
	[\bar{\qj}^{\qg}\qj^{\qg},\ql^{\qg},A_\mu^\qg]
	\exp\sum_{\qa=1}^{N_r}
        \intit\left[\intdxx \bar{\qj}^{\qa}[-\prt_{\qt}+iA^\qa_{\qt}+\mu
	-{\cal H}^\qa -V]\qj^{\qa}\right. \nn\\
	&& \left. 
	-\half\intd{^2 x d^2 x'}\ql^{\qa}(\vec x)
	U_0\inv(\vec x,\vec x')\ql^{\qa}(\vec x')
	+i\intdxx\ql^{\qa}\bar{\qj}^{\qa}\qj^{\qa}
	\right].
\eea
As a next step we perform a Fourier transform from imaginary time $\qt$ to
Matsubara frequencies. Since fermionic fields are antiperiodic on the
interval $[0,\qb]$, while bosonic fields are periodic, the allowed frequencies
for $\qj,\bar{\qj}$ and $A_{\qt},\ql$ are, respectively
\bea
	\qo_n=\fr{2\pi}{\qb}(n+\half) \;\;\;\mbox{(fermionic)}
	\hskip0.5cm & ; & \hskip0.5cm
	\nu_n=\fr{2\pi}{\qb}n\;\;\;
	 \mbox{(bosonic)}.
\eea
with $n$ integer.
We define the Fourier transformed fields by
\bea
	\qj^{\qa}(\qt)=\sum_{n=-\infty}^{\infty}\qj^{\qa}_n e^{-i\qo_n\qt}
	\hskip0.5cm &;& \hskip0.5cm
	\bar{\qj}^{\qa}(\qt)
	=\sum_{n=-\infty}^{\infty}\bar{\qj}^{\qa}_n e^{+i\qo_n\qt} \\
	\ql^{\qa}(\qt)=\sum_{n=-\infty}^{\infty}\ql^{\qa}_n e^{-i\nu_n\qt}
	 \hskip0.5cm &;& \hskip0.5cm 
	A_{\mu}^\qa(\qt)=\sum_{n=-\infty}^{\infty}(A_{\mu})^\qa_n
	e^{-i\nu_n\qt}
	\nn
\eea
which results in the following form of the action
\bea
	S &=& \qb\intdxx \qj\dagg(i\qo+i\hat{A}_{\qt}+\hat{\mu}+i\hat{\ql}
	-\hat{\cal H}-\hat{V})\qj \\ &&
	-\fr{\qb}{2}\intd{^2x d^2 x'}\ql\dagg(\vec x)
	U_0\inv(\vec x,\vec x')\ql(\vec x').
\eea

\begin{figure}
\begin{center}
\setlength{\unitlength}{1mm}
\begin{picture}(80,35)(0,0)
\put(0,0)
{\epsfxsize=3.5cm{\epsffile{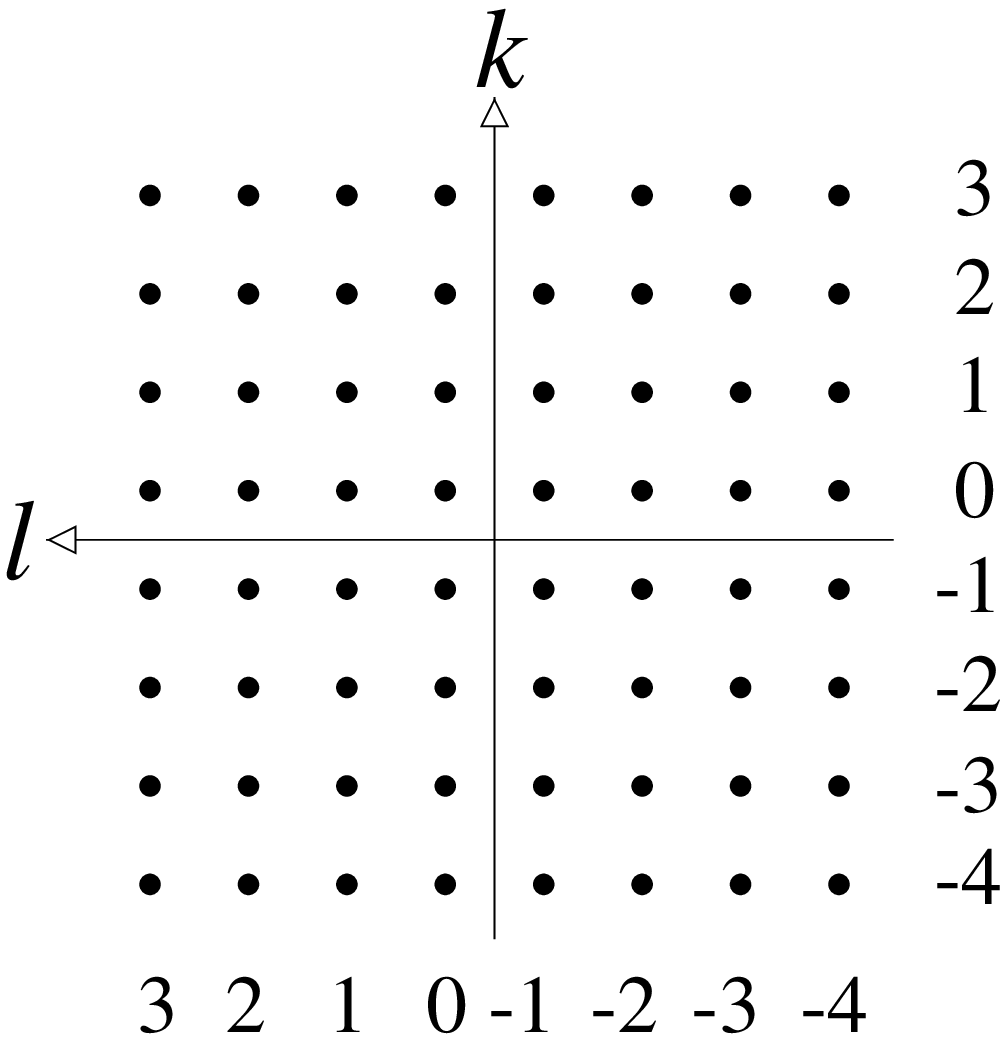}}}
\put(45,0)
{\epsfxsize=3.5cm{\epsffile{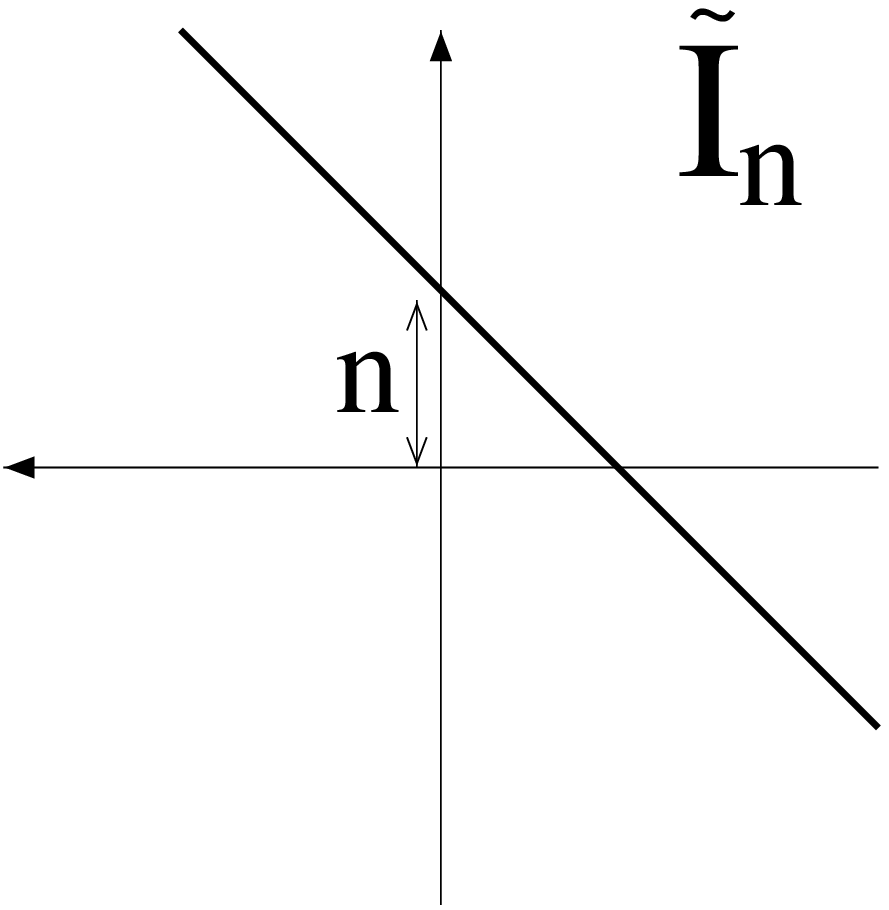}}}
\end{picture}
\caption{Our way of picturing a matrix  $[\cdots]_{kl}$ in
frequency space, and the structure of $\tI_n$ ($n\!>\!0$).}
\label{pictureI}
\end{center}
\end{figure}

\noindent
Here we have used matrix notation for combined replica and frequency
indices,
\be
	\qj\dagg(\cdots)\qj=\sum_{nm,\qa\qb}
	\bar{\qj}^{\qa}_n(\cdots)^{\qa\qb}_{nm}\qj^{\qb}_m.
\ee
The $\qo$ is a unity matrix in replica space, while in Matsubara space
it is a diagonal matrix containing the frequencies $\qo_n$
\be
	(\qo)^{\qa\qb}_{nm}=\qo_n\qd^{\qa\qb}\qd_{nm}.
\ee
The `hatted' quantities are defined according to
\be
	\hat{z}=\sum_{n,\qa}z^{\qa}_n \tI^{\qa}_n 
	\hskip1cm \mbox{ with }\;\;
	(\tI^{\qa}_n)^{\qb\qg}_{kl}=\qd^{\qa\qb}\qd^{\qa\qg}\qd_{k-l,n}.
\ee
The matrix $\tI^{\qa}_n$ is the unit matrix in the $\qa$'th replica
space, while in Matsubara space it is zero everywhere except on the
$n$'th diagonal, where it is 1.
The $\tI^{\qa}_n$-matrices are extremely important, because they will
turn out to be the
generators of the electromagnetic $U(1)$ transformations. 
But before we elaborate on this, let us first take the disorder
average of the replicated partition sum, in analogy with what has been
done in the free particle formalism.
This is done using a Gaussian distribution for the random potential 
$V(\vec x)$,
\bea
	\overline{Z} &=& \int\!{\cal D}[V] P[V] Z \\
	P[V] &\propto& \exp \left(-\fr{1}{2g}\intdxx V^2\right).\nn
\eea
This integration leads to a quartic term in the action of the form
$(\qj\dagg\qj)^2$, which can be decoupled by means of a
Hubbard-Stratonovich transformation, introducing
hermitian matrix field variables $\tQ^{\qa\qb}_{nm}(\vec x)$. 
The partition function now becomes
\bea
\label{ZpsiQ}
	Z &=& \int\!{\cal D}[\bar{\psi}\psi,\tQ,\ql, A_\mu]
	e^{S[\bar{\psi},\psi,\tQ,\ql,A_\mu]} \\
	S[\bar{\psi},\psi,\tQ,\ql,A_\mu] &=&
	-\fr{1}{2g}\Tr\tQ^2+\qb\intdxx
	\psi\dagg[i\qo+i\hat{A}_{\qt}+\hat{\mu}
	-\hat{\cal H}+i\hat{\ql}+i\tQ]\psi \nn\\
	&& -\fr{\qb}{2}\intd{^2x d^2 x'}\ql\dagg(\vec x)\ql(\vec x')
	U_0\inv(\vec x,\vec x')
\label{SpsiQ}
\eea
where the notation $\Tr$ stands for a trace over combined replica and
Matsubara indices as well as spatial integration $\intdxx$.
Notice that the only difference with the previously studied free particle
case \cite{a4} is that we work with a Matsubara frequency label,
rather than with advanced and retarded components alone.

\subsection{Gauge invariance; tunneling density of states}
\label{secgaugeinv}

A generic local $U(1)$ gauge transformation on the fermion fields and
the electromagnetic potentials has the form
\bea
\label{gaugetau}
	\qj^{\qa}(\vec x,\qt)\naar e^{i\chi^\qa(\vec x,\qt)}
	\qj^{\qa}(\vec x,\qt) \hskip0.5cm &;& \hskip0.5cm 
	\bar{\qj}^{\qa}(\vec x,\qt)\naar 
	e^{-i\chi^\qa(\vec x,\qt)}\bar{\qj}^{\qa}(\vec x,\qt) \\
	A_{\mu}^\qa(\vec x,\qt)\naar A_{\mu}^\qa(\vec x,\qt)
	+\prt_{\mu}\chi^\qa(\vec x,\qt)
	&&\nn
\eea
with $\chi^\qa$ real-valued functions periodic in $\qt$. In frequency
notation this gauge transformation is written as a unitary matrix
acting on the vector $\qj$
\bea
\label{gaugepsi}
	\qj\naar e^{i\hat\chi}\qj \hskip0.5cm &;& \hskip0.5cm \qj\dagg\naar\qj\dagg e^{-i\hat\chi}
	\\ 
	\hat{A}_{i}\naar\hat{A}_{i}+\prt_i \hat\chi \hskip0.5cm &;& \hskip0.5cm
	(A_\qt)^{\qa}_n\naar(A_\qt)^{\qa}_n-i\nu_n \chi^{\qa}_n.
\label{gaugeA}
\eea
From (\ref{gaugepsi}) it is clear that the $\tI$-matrices are the generators of
gauge transformations. 
From their multiplication
\be
	\tI^{\qa}_n\tI^{\qb}_m = \qd^{\qa\qb}\tI^{\qa}_{n+m}
\label{II=I}
\ee
it is readily seen that they span an abelian algebra, and that a gauge
transformation indeed acts in every replica channel separately, as seen in
(\ref{gaugetau}). 
The $\tQ$ transforms according to
\be
	\tQ\naar e^{i\hat\chi}\tQ e^{-i\hat\chi}.
\ee
The gauge invariance of the action (\ref{SpsiQ}) is easily checked:
First of all, the plasmon field $\ql$ and the combinations 
$\psi\dagg\tQ\psi$ and $\psi\dagg\psi$ are invariant. 
Secondly, the fact that the $\tI$ commute leads to 
$e^{-i\hat\chi}\nabla e^{i\hat\chi}=i\nabla\hat\chi$, 
from which it follows that the term $\psi\dagg\hat{\cal H}\psi$
is also invariant. Finally, using the following commutation relation,
\be
	[\tI^{\qa}_n,\qo]= -\nu_n \tI^{\qa}_n,
\ee
in combination with the transformation rule (\ref{gaugeA}) for $A_\qt$,
we find that the term 
$\psi\dagg(i\qo\! +\! i\hat{A}_\tau)\psi$ is also
invariant.
In the partition function (\ref{ZpsiQ}), the integration over fermion
fields can be performed, yielding an 
effective action for the variables $\tQ$, $A_\mu$ and $\ql$,
\bea
	S[\tQ,\ql,A_\mu] &=& -\fr{1}{2g}\Tr\tQ^2+\Tr\ln
	[i\qo+i\hat{A}_{\qt}+\hat{\mu}
	-\hat{\cal H}+i\hat{\ql}+i\tQ] \nn\\
	&& -\fr{\qb}{2}\intd{^2x d^2 x'}\ql\dagg(\vec x)\ql(\vec x')
	U_0\inv(\vec x,\vec x').
\label{SlambdaQA}
\eea
The gauge invariance of this effective action is again evident.
We only have to rewrite the gauge transformed second term 
$\Tr\ln[\cdots+ie^{i\hat\chi}\tQ e^{-i\hat\chi}]$ into
$\Tr\ln[e^{-i\hat\chi}(\cdots)e^{i\hat\chi}+i\tQ]$ and repeat the
arguments given above.

We end this section with an expression for the one particle Green's
function $G(\qt_2\! -\!\qt_1)$ which enters the tunneling density of
states,
\be
	G(\qt_2-\qt_1)=\langle\bar\qj^{\qa}(\vec x,\qt_2)
	\qj^{\qa}(\vec x,\qt_1)\rangle.
\ee
In terms of the $\tQ$-variable this expression is written as
\be
	\langle\tQ^{\qa\qa}(\qt_1,\qt_2)\rangle =
	\sum_{nm}e^{-i\qo_n\qt_1}e^{i\qo_m\qt_2}
	\langle\tQ^{\qa\qa}_{nm}\rangle.
\ee
The reader may verify that under a gauge transformation, the Green's
function transforms as
\be
	e^{i[\chi^\qa(\qt_2)-\chi^\qa(\qt_1)]}
	\langle\tQ^{\qa\qa}(\qt_1,\qt_2)\rangle
\ee
as it should.

\subsection{Truncation of frequency space}
\label{sectrunc}
\subsubsection{`Large' and `small' components}

We proceed as in the free particle analysis and split the $\tQ$ matrix
variable into `transverse' and `longitudinal' components,
\be
\label{QtoPT}
	\tQ=T\inv PT \hskip1cm  P=P\dagg \hskip1cm
	 T\in SU(2N').
\ee
Here, $P$ has only block-diagonal components in frequency space 
(i.e. $P^{\qa\qb}_{nm}$ is
nonzero only for $\qo_n\cdot\qo_m\! > \! 0$) and $T$ is a unitary
rotation. $2N'$ is 
the size of the Matsubara space times the number of replica channels, and
represents the size of the $\tQ$-matrix.

\begin{figure}
\begin{center}
\setlength{\unitlength}{1mm}
\begin{picture}(125,35)(0,0)
\put(0,0)
{\epsfxsize=25mm{\epsffile{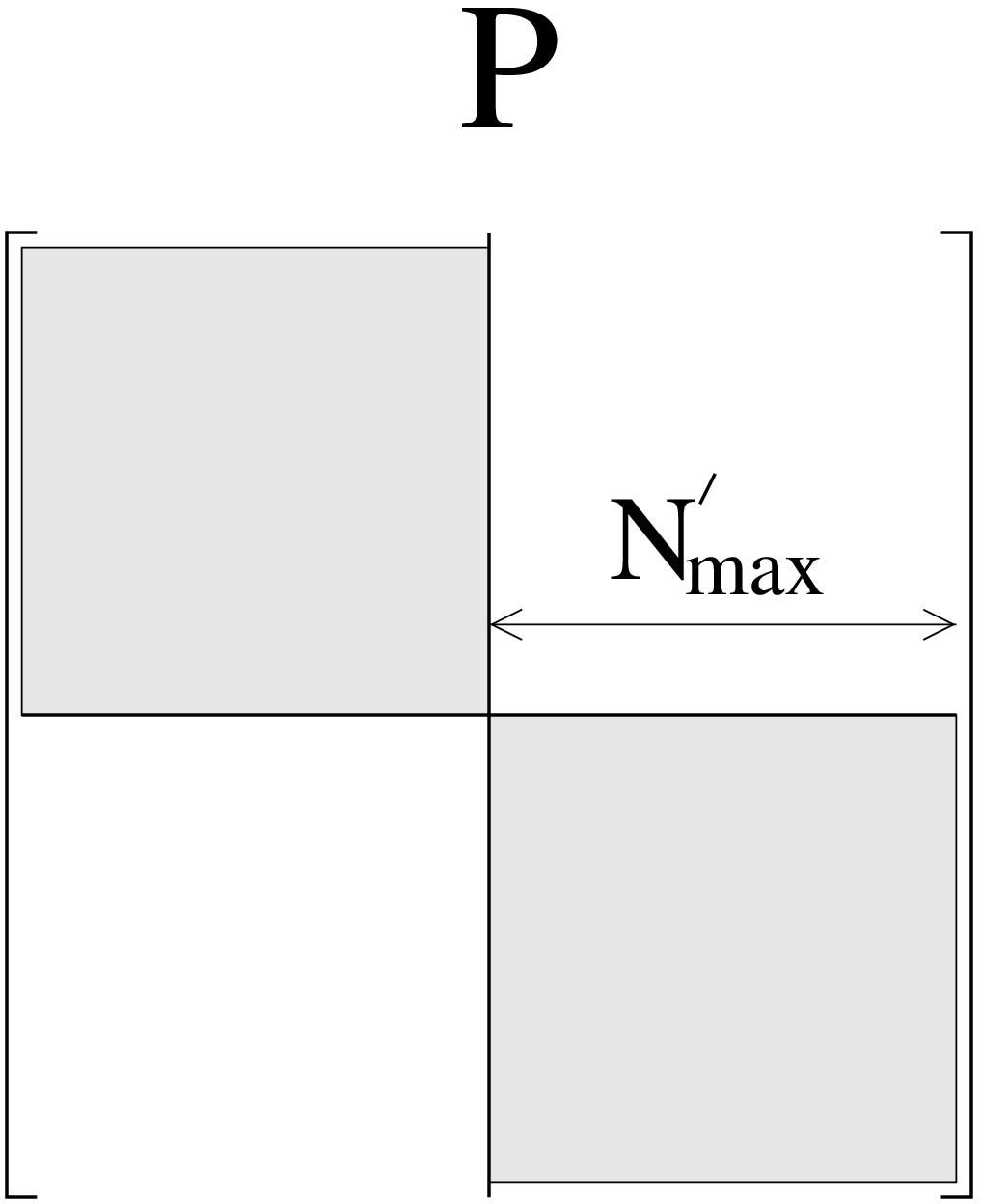}}}
\put(45,0)
{\epsfxsize=35mm{\epsffile{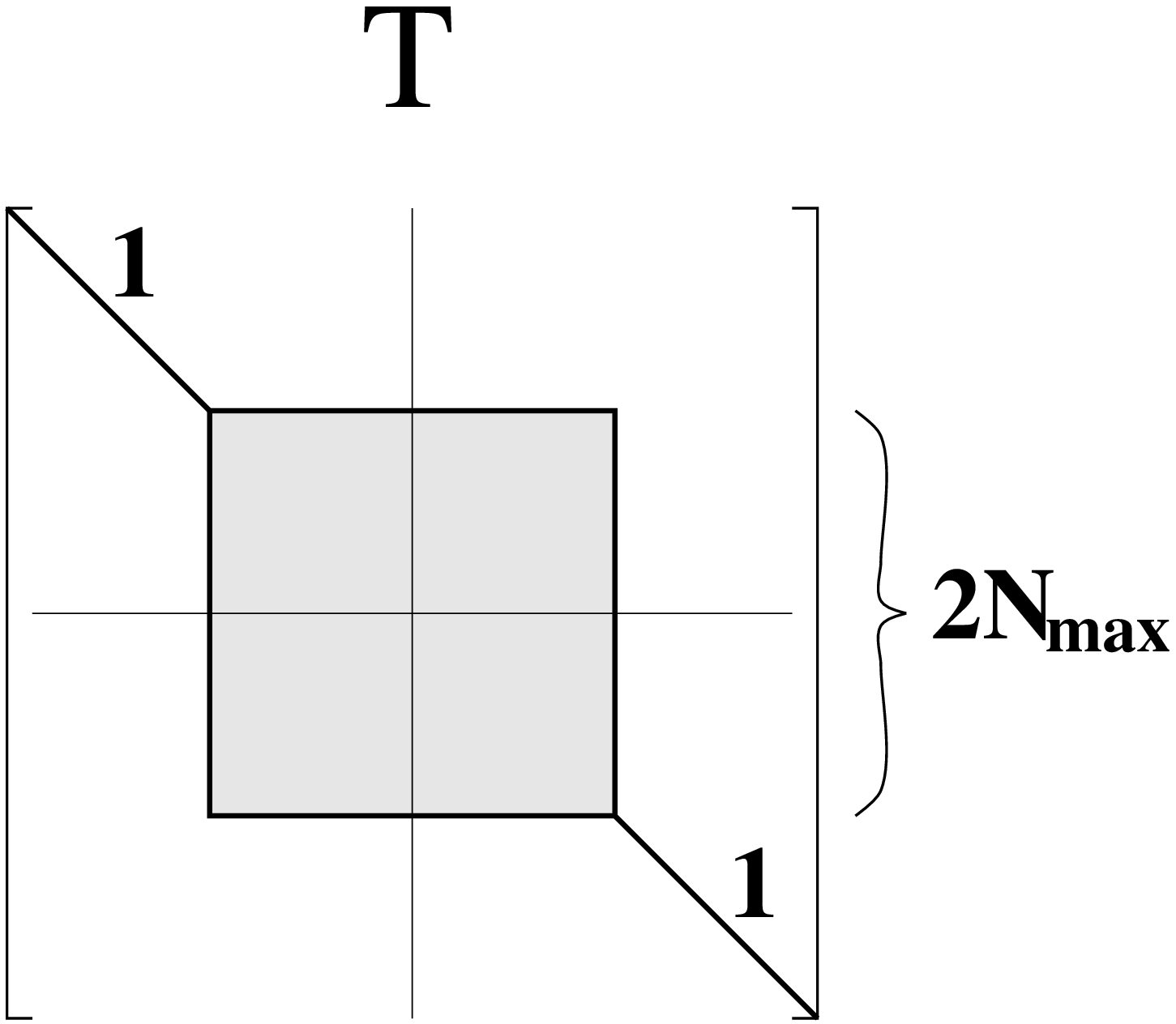}}}
\put(90,0)
{\epsfxsize=35mm{\epsffile{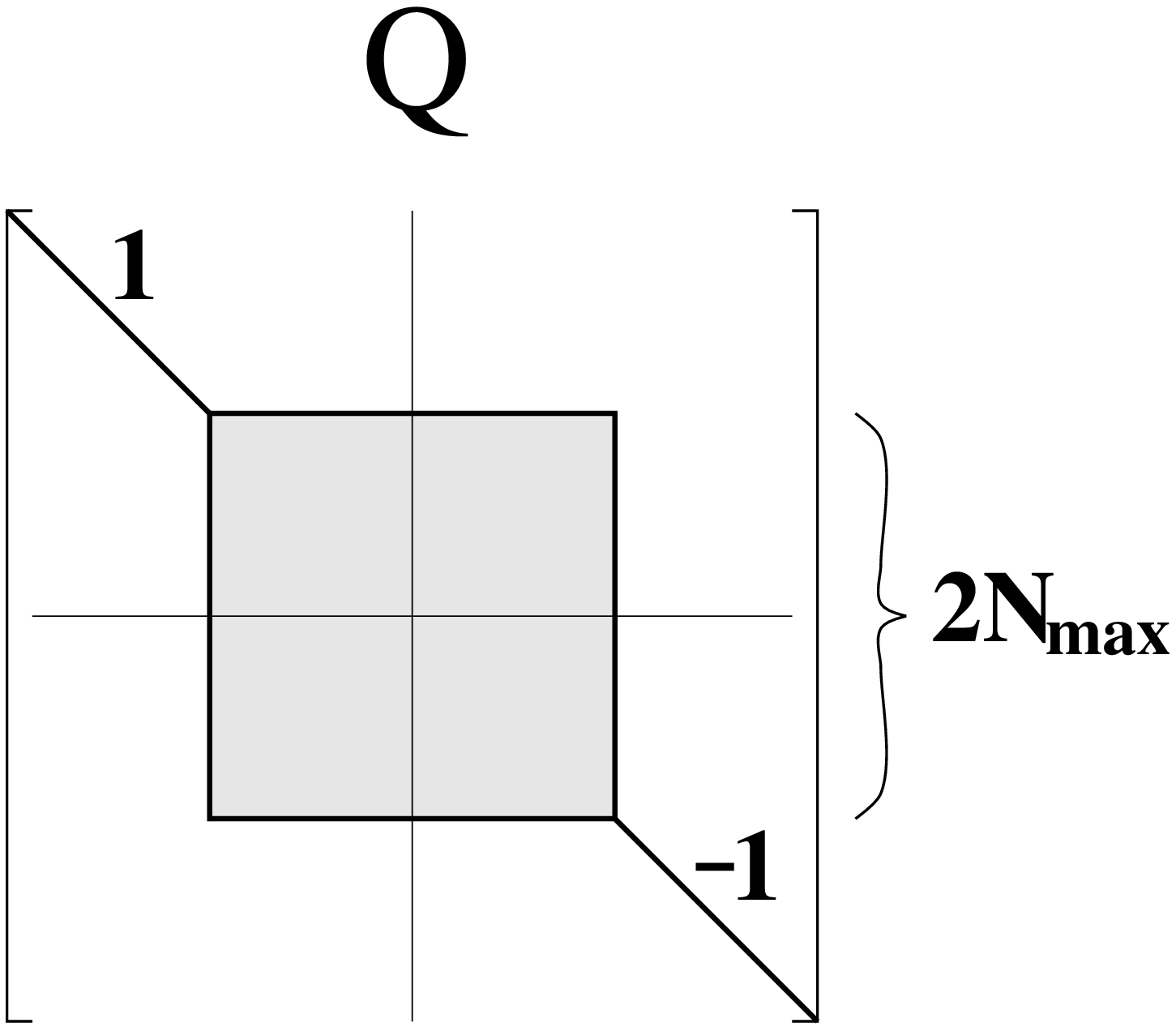}}}
\end{picture}
\caption{Sketch of the matrices $P$, $T$ and $Q$ in truncated
frequency space ($N_{\rm max}\!\ll\! N_{\rm max}'$).}
\label{figPTQ}
\end{center}
\end{figure}

This change of variables (\ref{QtoPT}) is motivated by the saddlepoint
structure of the theory (\ref{SlambdaQA}) in the absence of the fields
$A_{\mu},\ql$ and at zero temperature 
(i.e. $\qo_n\!\naar\! 0$). This saddlepoint
can be written as
\bea
\label{Qsp}
	\tQ_{\rm sp}\propto T\inv\qL T \hskip0.5cm &;& \hskip0.5cm
	\qL^{\qa\qb}_{nm}=\qd^{\qa\qb}\qd_{nm}\sgn(\qo_n),
\eea
indicating that the longitudinal fluctuations $P$ are the `massive'
components of the theory whereas the $T$-matrix fields are the lowest
energy excitations (Goldstone modes) in the problem. The manner in which
(\ref{QtoPT}) is going to be used is illustrated in 
Fig.~\ref{figPTQ}: we impose
on the $T$-rotations a cutoff $N_{\rm max}$ in Matsubara frequency
space, such that $1\ll N_{\rm max}\ll N_{\rm max}'$. 
It is important to keep in mind that working with a finite 
$N_{\rm max}$ is just a calculational device which will enable us to derive
an effective action for the lowest energy excitations $T$ by formally
integrating out the massive components $P$ (the latter can be done
explicitly by employing saddlepoint methods). Once an effective action for
the $T$-fields has been obtained, we have to find some procedure by which
the cutoff $N_{\rm max}$ can be sent to infinity. The main problem is to
ensure that such a procedure retains the electrodynamic $U(1)$ gauge
invariance of the theory. We will return to this problem at the end of
section~\ref{sectruncgaugeinv}. A more formal justification of the
`smallness' concept is postponed until section~\ref{secexternal} where we
introduce vector and scalar potentials in the effective action.

%The consideration of gauge invariance is of course a fundamental one.
%The manner in which it is handeled in the Finkelstein theory is
%completely new and hence, special care should be taken. We stress that
%this is precisely the reason why it is so important to go beyond the
%existing one-loop theory of localization and interaction effects and
%to prove the renormalizability of the theory to two-loop order. As it
%turns out in the remainder of this paper, this can be done only if the
%symmetries of the problem are respected by the renormalization group
%program that one chooses. Arbitrary renormalization schemes which do
%not respect these symmetries do not provide the information which is
%needed in order to decipher the consequences for the theory in the
%strong coupling limit where perturbation theory is no longer
%valid. Moreover, gauge non-invariant quantities such as the tunneling
%density of states give rise to severe infrared trouble, thus
%complicating the procedure of extracting the behaviour in the infrared
%for perturbative analysis. The latter, however, is a more general
%phenomenon, well recognised in gauge theories. All this says is that
%the above mentioned problem with gauge invariance, rather than being
%working hypothesis alone, actually has a much deeper significance in
%the theory as a whole and this just puts extra weight on the statement
%of renormalizability.

\subsubsection{${\cal F}$-invariance}

In order to be able to carry through the concept of `small' $T$-rotations
in a `large' Matsubara frequency space, we shall need to perform specific
algebraic manipulations which (sometimes) will be referred to by the name
of `${\cal F}$-algebra'.
To illustrate the meaning of this algebra we shall next derive the
effective action for the fields $T$ in the presence of Coulomb
interactions, but without scalar and vector potentials. The effective
action is defined by
\bea
	e^{S_{\rm eff}[T]} &\propto& \pathint{[P,\ql]}\exp\left(
	-\fr{1}{2g}\Tr P^2-\fr{\qb}{2}\intd{^2 x d^2 x'}
	\ql\dagg(\vec x)U_0\inv(\vec x, \vec x')\ql(\vec x') \right. \nn\\
	&& \left. +\Tr\ln[i\qo+\hat\mu+i\hat\ql-\hat{\cal H}+iT\inv PT]
	\right).
\label{defSeff}
\eea
In two spatial dimensions, the Coulomb interaction is infinitely
ranged. The Fourier transform is given by
\be
	U_0(\vec q)\propto \intdxx \fr{e^{-i\vec q\cdot\vec x}}{|\vec x|}
	\propto |\vec q|\inv.
\ee
From general symmetry considerations one can impose two important
conditions on the actual form of $S_{\rm eff}[T]$. 
\begin{enumerate}
\item
The only local
variable on which $S_{\rm eff}[T]$ can depend and which is consistent
with the symmetries of the problem is precisely of the form of 
$\tQ_{\rm sp}$ (\ref{Qsp}),
\be
\label{defQ}
	Q=T\inv\qL T.
\ee
Here the matrices are all acting in `large' frequency space of size
$2N_{\rm max}'\!\times\! 2N_{\rm max}'$. The $T$-rotations are effectively
`small' ($2N_{\rm max}\!\times\! 2N_{\rm max}$ with 
$N_{\rm max}\!\ll\! N_{\rm max}'$) as depicted in Fig.~\ref{figPTQ}. 
\item
The effective action must be invariant under global
(i.e. spatially independent) `$W$-rotations':
\bea
\label{SeffWrotinv}
	S_{\rm eff}[Q]=S_{\rm eff}[W_0 Q W_0\inv] 
	& \hskip5mm \mbox{with} \hskip5mm &
	W_0=\exp i\sum_{n,\qa}\chi^\qa_n \Ian,
\eea
where the matrix $\Ian$ stands for $\tI^\qa_n$ truncated to size
$2N_{\rm max}'\!\times\! 2N_{\rm max}'$.
\end{enumerate}
The statement (\ref{SeffWrotinv}), which is exact in the limit 
$N_{\rm max}' \!\!\naar\!\infty$, can easily be derived from (\ref{defSeff}) by
using the invariance of the Tr ln under unitary transformations. This
amounts to a spatially independent shift of the plasmon field $\ql$
inside the Tr ln according to
\be
	\ql^\qa_n(\vec x)\naar \ql^\qa_n(\vec x)+(\prt_\qt\chi)^\qa_n.
\ee
This shift can be absorbed in a redefinition of $\ql$ provided that the
interaction $U_0$ is infinitely ranged 
(i.e. $U_0\inv(\vec q)\!\naar\! 0$ as
$|\vec q|\!\naar\! 0$), as considered here.
The `${\cal F}$-invariance' of (\ref{SeffWrotinv}) plays a very special role in the
problem. Notice that (\ref{SeffWrotinv}) actually stands for a global
$U(1)$ gauge transformation and is directly related to the statement of
macroscopic charge conservation. The far-reaching consequences of this
statement were understood first by Finkelstein.

\subsubsection{Effective action}
We will proceed by presenting $S_{\rm eff}$ in an 
${\cal F}$-invariant manner as follows\cite{paperV}
\be
\label{SeffQ}
	S_{\rm eff}[Q] = S_\qs[Q]+S_{\rm F}[Q]+S_{\rm U}[Q].
\ee
The first term, $S_\qs$, is precisely the nonlinear $\qs$ model action
in the presence of the instanton term,
\be
	S_\qs[Q] = -\fr{1}{8}\qs^0_{xx}\Tr(\nabla Q)^2
	+\fr{1}{8}\qs^0_{xy}\Tr \qe_{ij}Q\prt_i Q\prt_j Q
\ee
where $\qs^0_{xx}$ and $\qs^0_{xy}$ denote the (mean field)
conductances in units $e^2/h$.
The second term, $S_{\rm F}$, can be written as
\be
\label{SF1}
	S_{\rm F}[Q] = \half z\fr{\pi}{\qb} 
	{\sum_{n\qa}}'\Tr[\Ian,Q][\Iamn,Q].
\ee
The quantity $z$ is the `singlet interaction amplitude'
and it defines the temperature scale. The prime on
the summation over $n$ indicates a restriction on the frequency range,
$n\!\in\!\{-N_{\rm max}',\cdots,N_{\rm max}'\!-\!1 \}$.
The last term in (\ref{SeffQ}) contains the Coulomb interaction $U_0$
explicitly and can be written as 
\be
\label{SU}
	S_{\rm U}[Q] = -\fr{\pi}{\qb}{\sum_{n\qa}}\intd{^2 x d^2 x' }
	[\tr \Ian Q(\vec x)]U\inv(\vec x-\vec x')
	[\tr \Iamn Q(\vec x')].
\ee
In momentum space $U\inv$ is given by
\be
\label{defU}
	U\inv(p)=\int\! \frac{d^2 r}{2\pi}U\inv(\vec r)e^{-i\vec p\cdot\vec r}
	=\frac{\pi}{2}\cdot\frac{1}{\qr\inv+U_0(p)}
\ee
where $\qr\!=\!\prt n/\prt\mu$ is the {\em thermodynamic density of
states}. 

\subsubsection{Examples of ${\cal F}$-algebra}
We stress again that from now onward, all matrix manipulations are done in
truncated ($2N_{\rm max}'\!\times 2N_{\rm max}'$) 
Matsubara frequency space. The truncated I-matrices
obviously no longer obey the simple $U(1)$ algebra
(\ref{II=I}), but instead
\bea
	(\Ian I^\qb_m)^{\mu\nu}_{kl}=(\tI^\qa_n \tI^\qb_m)^{\mu\nu}_{kl}
	g_{l+m} \hskip0.5cm &;& \hskip0.5cm
	[\Ian, {\rm I}^\qb_m]^{\mu\nu}_{kl}=\qd^{\qa\qb\mu\nu}\qd_{k-l,m+n}
	(g_{l+m}-g_{l+n})
\eea 
where $\qd^{\qa\qb\mu\nu}$ means that all replica indices have to be the
same, and $g_i$ is a step function equal to one if 
$i\in\{ -N_{\rm max}',\ldots,N_{\rm max}' \! -\! 1   \}$ and zero otherwise.
Consequently, the $W_0$ in (\ref{SeffWrotinv})
stands for a more complicated unitary matrix of size
$2N_{\rm max}'\!\times\! 2N_{\rm max}'$. Nevertheless, by making use
of elementary but subtle algebra one can show that the procedure with
an arbitrary `large' cutoff correctly describes the low energy sector
and correctly retains the electrodynamic gauge invariance of the
theory at low frequencies. We proceed by listing some
of the important subtleties of the ${\cal F}$-algebra.

\begin{figure}
\begin{center}
\setlength{\unitlength}{1mm}
\begin{picture}(35,45)(0,0)
\put(0,5)
{\epsfxsize=35mm{\epsffile{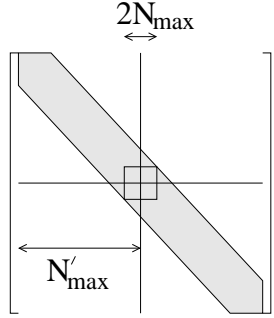}}}
\end{picture}
\caption{The summation interval 
$n\in\{ -2N_{\rm max}\! +\! 1,\cdots,2N_{\rm max}\! -\! 1  \}$,
indicated by the shaded area.}
\label{figIrange}
\end{center}
\end{figure}

\noindent
The definition of (\ref{SF1}), which involves the restricted frequency
sum, is particularly delicate. It can be written in a more familiar
form by first writing
\be
	Q=\qL+\qd Q.
\ee
Written out explicitly, (\ref{SF1}) now becomes
\be
	S_{\rm F}[Q]=z\fr{\pi}{\qb}\intdxx \left[ \sum_\qa
	\sum_{klmn}\qd Q^{\qa\qa}_{kl}\qd Q^{\qa\qa}_{mn}\qd_{k-l,n-m}
	+4\tr\qh\qd Q \right] +\mbox{const.}
\label{SF3}
\ee
which is the result originally obtained by Finkelstein.
Here $\qh$ is a matrix representation of the Matsubara frequencies,
\be
	\qh^{\qa\qb}_{nm}=n \qd^{\qa\qb}\qd_{nm}.
\ee
The constant in (\ref{SF3}) is proportional to 
$\tr\qh\qL\!=\!N_r\sum_{n=-L}^{L-1} |n|$, where the cutoff $L$ depends
on the exact definition of the prime in (\ref{SF1}). 
Mostly, we will not be interested in the exact value of $L$, and a
prime on a summation simply means a restriction to small frequencies.
Using the definition of the prime under (\ref{SF1}),
Eq. (\ref{SF3}) can be re-expressed in terms of $Q$ as follows
\be
	S_{\rm F}[Q]=z\fr{\pi}{\qb}
	\intdxx \left[ {\sum_{\qa n}}(\tr \Ian Q)
	(\tr \Iamn Q)+4\tr\qh Q-6\tr\qh\qL
	\right].
\label{SF2}
\ee
Notice that the bilinear forms in $Q$ in (\ref{SF1}) and (\ref{SF2})
differ by a frequency term $\tr\qh Q$ and a constant.
Within Finkelstein's formulation of the problem (\ref{SF3}), the very
special relative coefficient ``4'' between the first (singlet
interaction) and second (frequency) term arises from the macroscopic
conservation laws in a very indirect and deep manner. The advantage of
the present formalism is (amongst other things) the simple algebraic
interpretation of the result which can be obtained from the symmetries
of the problem. Moreover, the constant appearing in (\ref{SF2}) has a
very special significance for physical quantities such as the specific
heat. This aspect of the problem will be further discussed in
subsequent work\cite{b33}.

\subsubsection{General remarks}

For a general understanding of the result (\ref{SeffQ}) we next discuss the
various pieces separately. First, by putting the temperature $\qb\inv$
equal to zero we obtain the same result $S_\qs$ as in the free
electron theory. The `bare' parameters $\qs^0_{xx}$ and $\qs^0_{xy}$
are generally modified by the electron-electron interactions. The
modifications are of a Fermi-liquid type and in the limit of strong
magnetic fields the results depend on the ratio of disorder energy
$\qG_0$ (the width of the Landau band) and the typical Coulomb energy
$E_0$ ($=U_0(\ell)$, where $\ell$ is the magnetic length). 

The most important part next is $S_{\rm F}$. Quite unlike what one
naively might expect, the presence of $S_{\rm F}$ alters the
ultraviolet singularity structure of the free electron theory $S_\qs$
altogether\cite{b33}. This peculiar aspect of the problem
indicates that the electronic system with Coulomb interactions has a behavior
fundamentally different from that with finite range interactions or
free electrons.

Next we briefly elaborate on the significance of the Coulomb term
$S_{\rm U}$ (\ref{SU},\ref{defU}) which is usually ignored in
renormalization group analyses, since it really stands for a higher
dimensional operator (notice that $U\inv(p)\propto |p|$ in the small
momentum limit). The importance of this term, however, can not be
overemphasized. First, we mention that in the large momentum limit we
can substitute $\fr{\pi}{2}\qr$ for $z$ in (\ref{SF1}) \cite{paperV}.
In this limit the sum of $S_{\rm U}$ and
$S_{\rm F}$ does not contain the term quadratic in (tr I$Q$) and
(\ref{SeffQ}) reduces to the effective action for {\em free}
particles. This means that the full theory of (\ref{SeffQ}-\ref{SU})
is appropriately interpreted in terms of a cross-over phenomenon
between free electron behaviour at short distances (or high
temperature $\qb\inv$) and an interaction dominated behaviour which
appears at larger distances (or lower temperatures $\qb\inv$) only.

Secondly, the complete form of (\ref{SU}) and (\ref{defU}) unravels
important information on the nature of quantum transport even for
ordinary metals. This will be the main subject of
section~\ref{secresponsetree}, where we compute at a tree level the
complete momentum and frequency dependent response of the theory to
electromagnetic perturbations. 

Finally, we mention that although (\ref{SU}) is naively irrelevant from the
weak coupling renormalization point of view, it nevertheless dominates
the quantum transport problem in the strong coupling (insulating)
phase which is characterized by strong interaction effects such as the
appearance of the Coulomb gap in the (quasiparticle) density of
states.
This will be the main subject of a subsequent paper, where we embark
on the renormalization group behaviour of the theory\cite{b33}.

\subsection{Gauge invariance in truncated frequency space}
\label{sectruncgaugeinv}

In this section we wish to show that $S_{\rm eff}$ (\ref{SeffQ}) is
${\cal F}$-invariant, i.e. it satisfies the requirement stated in
(\ref{SeffWrotinv}). 
We will proceed by giving the results for an arbitrary spatially 
{\em dependent}
gauge or W-transformation from which the statement of 
${\cal F}$-invariance follows
trivially. Assuming that $W$ approaches unity at spatial infinity, we
obtain
\be
	S_\qs[WQW\inv]=-\fr{1}{8}\qs^0_{xx}\Tr (\vec d Q)^2
	+\fr{1}{8}\qs^0_{xy}\Tr \qe_{ij}Q\prt_i Q\prt_j Q
\label{transSsig}
\ee
where  
\be
	\vec d Q=[\nabla+i\nabla\hat\chi,Q].
\ee
Furthermore we have \cite{foot1}
\bea
	S_{\rm F}[WQW\inv] &=& S_{\rm F}[Q] \\
	S_{\rm U}[WQW\inv] &=& -\fr{\pi}{\qb}{\sum_{\qa n}}
	\intd{^2 x d^2 x'}\left[
	\tr \Ian Q(\vec x)+\fr{\qb}{\pi}(\prt_\qt \chi)^\qa_{-n}(\vec x)
	\right]  U\inv(\vec x,\vec x') \nn\\ && \times
	\left[\tr \Iamn Q(\vec x')+
	\fr{\qb}{\pi}(\prt_\qt \chi)^\qa_n(\vec x')\right].
\eea
For completeness we give the results for the W-transformations of 
$\tr \Ian Q$ and $\tr\qh Q$:
\bea
\label{transIQ}
	\tr \Ian WQW\inv &=&  \tr \Ian Q+\fr{\qb}{\pi}
	(\prt_\qt \chi)^\qa_{-n} \\
	\tr\qh WQW\inv &=& \tr\qh Q-\fr{\qb}{2\pi}\tr Q\widehat{\prt_\qt\chi}
	-(\fr{\qb}{2\pi})^2(\prt_\qt\chi)\dagg(\prt_\qt\chi).
\label{transetaQ}
\eea
The remarkable aspect of these results (\ref{transSsig}-\ref{transetaQ}) is
that the W-rotation on the $Q$ does not contribute beyond the lowest few orders
in a power series expansion in the I-matrices! What is more, the arbitrary
cutoff $N_{\rm max}'$ does not enter these final results and can be safely
taken to infinity. Next, from (\ref{transetaQ}) we see that within the
$Q$-field formalism the frequency matrix $\qo$ does not transform
simply according to the linear rule 
$\qo\!\naar\!\qo\! -\!\widehat{\prt_\qt\chi}$
as one would naively expect. The consistency of the 
${\cal F}$-algebra demands
that terms quadratic in the gauge field $\chi$ are being generated
such that the Finkelstein part of the action ($S_{\rm F}$) as a whole
remains gauge invariant. The results of the ${\cal F}$-algebra are therefore
somewhat counterintuitive.

In summary we can say that electrodynamic gauge transformations 
can be
incorporated in the $Q$-field theory for localization and interaction
effects. For this purpose we introduced the `smallness' concept for
the $Q$-fields, whereas the W or electrodynamic gauge transformations
are considered to be `large'. 
In the next section (\ref{secexternal}) we will build upon these findings
and present a formal justification of our cutoff procedure in
Matsubara frequency space. In practice this means that the coupling
between the `large' $W$-rotations and the `small' $Q$-matrix fields as
discussed in this section is the {\em only consistent way} of carrying
through electrodynamic gauge transformations in the effective action
formalism. The stringent requirements on the cutoff procedure do not,
however, provide an answer to the fundamental question of $U(1)$ gauge
{\em invariance}. More specifically, since the $Q$ and $WQW\inv$ do
not (by construction) belong to the same manifold, we generally can
not absorb the $W$-rotation into the measure of the $Q$-integration
and prove the gauge invariance in this way. The general idea behind
this approach, however, is that gauge invariance is only obtained
after the cutoff $N_{\rm max}$ in the effective action is sent to
infinity. This way of handling the $U(1)$ gauge invariance is
completely new and special care should therefore be taken. The proof
of gauge invariance of the Finkelstein theory ultimately relies on the
results of explicit, laborious calculations, both perturbative
\cite{b33} and non-perturbative \cite{PruiskenBaranov}. This, then, puts extra
weight on statements of renormalizability and we will embark on this
problem in subsequent papers.

From now onward we are going to treat the $W$-rotations and ${\cal
F}$-invariance as a good symmetry of the problem, keeping in mind that
the limit $N_{\rm max}\!\!\naar\!\infty$ is always taken in the end.

\subsection{External fields}
\label{secexternal}

One may next employ the results of the previous section and extend the
theory by including vector and scalar potentials $A_\mu$. 
This could be done in such a way that the resulting action is
invariant under the transformation 
$Q\!\naar\! e^{i\hat\chi}Qe^{-i\hat\chi}$, 
$A_\mu^\qa\!\naar\! A_\mu^\qa\! +\!\prt_\mu\chi^\qa$. Such a procedure,
however, does not imply anything for the topological piece of the
action $S_\qs$, which couples to external fields in a more
complicated fashion. In anticipation of a detailed analysis of
disordered edge currents \cite{b31} we report the following
results. 
\bea
\label{SsigwithA}
	S_\qs &\naar& -\fr{1}{8}\qs_{xx}^0\Tr([\vec D,Q])^2
	+\fr{1}{8}\qs_{xy}^0\qe_{ij}\Tr Q[D_i,Q][D_j,Q] 
%	-\fr{i}{2}\qs_{xy}^{II}\sum_{n\qa}\intdxx [\tr\Ian Q
%	-\fr{\qb}{\pi}(\tilde A_\qt)^\qa_{-n}]B^\qa_n
	-\fr{\qb}{8\pi^2\qr}(\qs_{xy}^{II})^2\intdxx B\dagg B \\
	S_{\rm U} &\naar& -\fr{\pi}{\qb}{\sum_{n\qa}}\int\!
	\fr{d^2 q}{(2\pi)^2} U\inv(\vec q)
	\left[
	\tr \Ian Q(-\vec q)-\fr{\qb}{\pi}(\tilde A_\qt)^\qa_{-n}(-\vec q)
	\right]
	\left[
	\tr \Iamn Q(\vec q)-\fr{\qb}{\pi}(\tilde A_\qt)^\qa_n(\vec q)
	\right]
\label{SUwithA}
\eea
where we have defined
\be
	\tilde A_\qt=A_\qt-\fr{i}{2\pi\qr}\qs_{xy}^{II}B.
\ee
The terms containing 
$\qs_{xy}^{II}\!\propto\! \prt n/\prt B$ \cite{a6,c3} are the result of
the diamagnetic edge currents in the problem, which give rise to extra
contributions. We stress that the complete microscopic result of
(\ref{SsigwithA}) clearly demonstrates the theoretical subtleties of
the effective action procedure which can not be taken for granted. In
addition to this, we mention that (\ref{SsigwithA}) and
(\ref{SUwithA}) really stand for extremely nontrivial statements made
on the low-energy, long wavelength excitations of the theory.
In order to see this, we consider the theory (\ref{SsigwithA},
\ref{SUwithA}) at a classical level, i.e. we put $Q\! =\!\qL$. The
results now represent an effective action for the external fields
$A_\mu$ which contains the {\em same} microscopic parameters
$\qs_{ij}^0$ etc. as those appearing in $S_{\rm eff}[Q]$
(\ref{SeffQ}-\ref{SU}) without external fields. This result is truly
remarkable if one realizes that the effective actions $S_{\rm
eff}[A_\mu, Q\! =\!\qL]$ and $S_{\rm eff}[A_\mu\! =\! 0, Q]$ follow from
fundamentally different expansion procedures applied to the original
theory (\ref{SlambdaQA}-\ref{QtoPT}). In appendix A we elaborate
further on this point and show that the different expansion procedures
are in fact related by Ward identities. These Ward identities are not only
crucially important in the microscopic derivation of the general
result (\ref{SsigwithA}, \ref{SUwithA}), they also provide a formal
justification of the `smallness' concept. In appendix B we give a
simple example and show how the theory (\ref{SeffQ}-\ref{SU}) can be
obtained in this way.

\ns{Response at tree level}
\label{secresponsetree}
\subsection{Perturbative expansion}

It is straightforward to check that the $Q$-field theory at a
classical level (putting $Q\! =\! \qL$) does not provide a gauge invariant
response to the external fields $A_\mu$. In order to obtain a $U(1)$
invariant result, one has to work with the propagators of the
$Q$-field fluctuations. A $U(1)$ invariant result at a so-called tree
level is obtained by taking the $Q$-field fluctuations to lowest order
into account. 

The most effective way to proceed is to first make use of a $W$- or
gauge transformation such that the $A_\qt$ in (\ref{SUwithA}) is absorbed
into the vector potential $\vec A$. It is easy to check that under
such a $W$-rotation the fields transform according to
\bea
	\vec A^\qa_n &\naar&  z^\qa_n =
	\vec A^\qa_n+\fr{\nabla (A_\qt)^\qa_n}{i\nu_n} \\
	A_\qt &\naar& 0. \nn
\eea
It is obviously advantageous to deal directly with the gauge invariant
quantity $\vec z^\qa_n\! =\!i\vec E^\qa_n/\nu_n$, where $\vec E$ is the
electric field $(\prt_\qt\vec A\! -\!\nabla A_\qt)$.
In order to define a perturbative expansion in the $Q$-field we write
\be
	Q=\left(\matrix{ \sqrt{1-qq\dagg} & q \cr
	q\dagg & -\sqrt{1-q\dagg q}}
	\right)
\ee
where the matrices $q,q\dagg$ are taken as independent field
variables.
We use the following convention for the Matsubara indices: the
quantities $n_1,n_3,\cdots$ with odd subscripts run over non-negative
integers, such that the corresponding fermionic frequencies
$\qo_{n_i}$ are positive. By the same token, the $n_2,n_4,\cdots$ run
over negative integers and the corresponding $\qo_{n_i}$ are all
negative. 
The action can be written as a series in powers of the fluctuation
fields $q,q\dagg$. The propagators of the Gaussian theory are given by
\bea
	\left\langle q^{\qa\qb}_{n_1 n_2}(p)\;\;
	[q\dagg]^{\qg\qd}_{n_4 n_3}(-p')
	\vphantom{H^H}\right\rangle &=&
	\fr{4}{\qs_{xx}^0}\qd^{\qa\qd}\qd^{\qb\qg}\qd(p-p')
	\qd_{n_{12},n_{34}} D_p(n_{12}) \times \nn\\
	&& \times \left\{\qd_{n_1 n_3}
	+\qd^{\qa\qb}\qk^2[z-U\inv(p)]
	D_p^c(n_{12})\right\}
\eea
\bea
	D_p(m)=\left[p^2+\qk^2 mz\right]\inv
	\hskip0.5cm &;& \hskip0.5cm
	D_p^c(m)=\left[p^2+\qk^2 mU\inv(p)\right]\inv \nn\\
	\qk^2=\fr{8\pi}{\qb\qs_{xx}^0} \hskip0.5cm &;& \hskip0.5cm
	n_{12}=n_1-n_2. \nn
\eea
We obtain the following result for the response at tree level
\be
	S[A_\mu]=-\qs_{xx}^0\sum_{\qa,n>0}\int\!
	\frac{d^2 p}{(2\pi)^2}
	n\;\;\overline{(z_i)^\qa_n(p)}\left[\qd_{ij}
	-\frac{p_i p_j}{p^2+\qk^2 nU\inv(p)}
	\right](z_j)^\qa_n(p)
\label{responsexx}
\ee
where, for simplicity, we have put $\qs_{xy}^0\! =\! 0$ for the moment. (The
bar-notation indicates complex conjugation.)
The theory (\ref{responsexx}) provides important physical information
on the process of quantum transport. In order to show this we write
for the electron density $n$ (using $\qt\! =\! it$)
\be
	-\qb n^\qa_m(p)=\frac{\qd S[A_\mu]}{\qd (A_\qt)^\qa_{-m}(-p)}
	= -\fr{\qb\qs_{xx}^0}{2\pi} p_i\left[\qd_{ij}
	-\frac{p_i p_j}{p^2+\qk^2 mU\inv(p)}
	\right](z_j)^\qa_m(p)
\label{densderiv}
\ee
which can be written as
\be
	[\nu_m+\fr{1}{4}\qs_{xx}^0 p^2 U(p)]n^\qa_m
	=\fr{\qs_{xx}^0}{\qb}m\vec p\cdot
	\vec z^\qa_m=
	i\vec p\cdot (\vec\jmath_{\rm ext})^\qa_m.
\label{density}
\ee
We have obtained a current density on the r.h.s. by using
$\vec z^\qa_m\! =\! i\vec E^\qa_m/\nu_m$ and 
$\vec\jmath_{\rm ext}\! =\! \fr{\qs_{xx}^0}{2\pi}\vec E$. 

Eq.~(\ref{density}) can be rewritten in the form
\be
	[\nu_m+p^2 D_{xx}^0](n_c)^\qa_m+i\fr{\qs_{xx}^0}{2\pi}
	\vec p\cdot[\vec E-i\vec p U_0 n_c]^\qa_m=0,
\label{dens2}
\ee
where $n_c\!=\!-n$ is the charge density and $D_{xx}^0$ the diffusion
constant, equal to $\qs_{xx}^0/2\pi\qr$ by the Einstein relation. In
spacetime notation (\ref{dens2}) reads
\be
	\prt_t n_c +\nabla\cdot(\vec\jmath_{\rm diff}+\vec\jmath_c)=0
\label{konzerva}
\ee
and expresses the well known result from the theory of
metals \cite{Nozieres} with
$\vec\jmath_{\rm diff}\!=\! -D_{xx}^0 \nabla n_c$ being the diffusive
current and $\vec\jmath_c\!=\!\fr{\qs_{xx}^0}{2\pi}\vec E_{\rm tot}$
the conductivity current generated by the total electric field inside
the system:
\be
	\vec E_{\rm tot}=\vec E-\nabla\intd{^2 x'}U_0(x-x')n_c(x').
\ee
Notice that in the limit of low momenta (or high frequencies) the 
$\vec\jmath_{\rm diff}$ in (\ref{konzerva}) can be neglected and the
system only responds to the sum of externally applied and internally
generated electric fields. The instantaneous Coulomb potential
apparently wins over the much slower diffusive processes in this
case. In a separate paper\cite{paperV} we address the problem of quantum
corrections to the semiclassical theory (\ref{konzerva}).

\subsection{Including magnetic fields}
\label{secmagnetic}

The general result (\ref{SsigwithA}) describes interesting edge
dynamics in case strong magnetic fields are present \cite{b31}. In the
remainder of this paper, however, we will limit ourselves to the
problem of weak magnetic fields, in which case the $\qs_{xy}^{II}$
term can be neglected \cite{c3} and edge effects become
immaterial. The topological piece of (\ref{SsigwithA}) can then be
written as
\be
	\fr{1}{4}\Tr \qe^{ij}Q[D_i,Q][D_j,Q]=\fr{1}{4}\Tr
	\qe^{ij}Q\prt_i Q \prt_j Q+i\Tr Q\curl\hat{\vec z}
	+\intdxx\sum_{\qa,n}n\vec z^\qa_n\!\times\! \vec z^\qa_{-n}.
\ee
This leads to the following gauge invariant response
\bea
\label{responsexy}
	S[A_\mu] &=& -\sum_{\qa,n>0}
	\int\!\fr{d^2 p}{(2\pi)^2}n\;\overline{(z_i)^\qa_n(p)}
	\left[\qs_{xx}^0\qd_{ij}+\qs_{xy}^0\qe_{ij}\right](z_j)^\qa_n(p)
	\\ && +
	\fr{1}{\qs_{xx}^0}\!\!\sum_{\qa,n>0}\int\!\fr{d^2 p}{(2\pi)^2}n\;
	\left[\qs_{xx}^0\vec p\cdot\!\vec z^\qa_n+\qs_{xy}^0\vec p\times\!\vec
	z^\qa_n\right]^* D_p^c(n)
	\left[\qs_{xx}^0\vec p\cdot\!\vec z^\qa_n-\qs_{xy}^0\vec p\times\!\vec
	z^\qa_n\right]. \nn
\eea
If we now repeat the calculation of the electron density in
section~\ref{secresponsetree} using the action (\ref{responsexy}), we
find that the results (\ref{density})-(\ref{konzerva}) still hold,
with one modification: The `external' current $\vec\jmath_{\rm ext}$
and the internally generated current $\vec\jmath_c$
now also include a Hall current,
\be
	j_i=\fr{\qs_{xx}^0}{2\pi}E_i 
	+\fr{\qs_{xy}^0}{2\pi}\qe_{ij}E_j.
\ee
(The modification of $\vec\jmath_c$ is not apparent in the calculations,
however, since 
$\nabla\cdot\vec{\jmath}_c^{\;\rm Hall}\!\propto\! 
\qe_{ij}q_i q_j U(q)n(q)\! =\! 0$.)
For convenience later on, we write the result (\ref{responsexy}) in
terms of new variables $\qF,\qJ$ 
\be
	z_i=\prt_i\qF+\qe_{ij}\prt_j\qJ
\ee
\be
	S[\qF,\qJ]=-\qs_{xx}^0\sum_{\qa,n>0}\int\!\fr{d^2 p}{(2\pi)^2}
	n p^2
	(\qF^*, \qJ^*) M \left(\matrix{\qF\cr\qJ}\right).
\label{SPhiPsiform}
\ee
The $2\!\times\! 2$ matrix $M$ is given by
\bea
	M=\left[\matrix{{\cal G}& -\qo_c\qt{\cal G} \cr
	\qo_c\qt{\cal G} & 1+(\qo_c\qt)^2(1-{\cal G})}
	\right] 
	& \hskip5mm \mbox{ with }\hskip5mm &
	{\cal G}=1-p^2 D_p^c(n)
\label{defM}
\eea
where we have made use of the semiclassical notation
\bea
	\qs_{xx}^0=\frac{\qs_0}{1+(\qo_c\qt)^2} \hskip0.5cm &;& \hskip0.5cm
	\qs_{xy}^0=\qo_c\qt\;\qs_{xx}^0.
\label{sigsemiclass}
\eea

\ns{Chern-Simons gauge fields}
\label{secCS}
\subsection{Introducing CS gauge fields}
\label{secintroCS}

The results of the previous sections are easily extended to include
statistical gauge fields and the Chern-Simons action, leading to the
composite fermion description of the half-integer effect in the
quantum Hall regime. The action (\ref{basicaction}) now becomes
\be
	S[\bar\psi,\psi,A_\mu]\naar S[\bar\psi,\psi,A_\mu+a_\mu]
	+\fr{i}{4\pi}\fr{1}{2p}\int\! a\wedge da
\label{2pfluxes}
\ee
where we have used the shorthand notation
\be
	\int\! a\wedge da=\int_0^\qb\! d\qt\intdxx \qe^{\mu\nu\ql}
	a_\mu\prt_\nu a_\ql.
\ee
Equation (\ref{2pfluxes}) describes the coupling of an even number
($2p$) of flux quanta to each electron, but it leaves the physical
amplitudes of the theory formally unchanged. This flux binding
transformation has been exploited at many places elsewhere and it
leads to the composite fermion description of the qHe. The action
(\ref{2pfluxes}) can be directly translated into $Q$-field theory 
by writing $A_\mu\!\naar\! A_\mu\! +\! a_\mu$
with
one important subtlety, namely that the zero-frequency components of
$a_\mu$ obviously commute with the $T$-rotations and hence belong to
the underlying theory with the $P$ matrix field. These zero-frequency
components can be treated in mean-field theory. Writing
$b\! =\!\curl\vec a$, the mean field equation for the Chern-Simons
magnetic field is
\be
	b=-2p\;n(B+b)
\label{CSb}
\ee
where the composite fermion density $n(B+b)$ is defined by
\be
	n(B+b)=L^{-2}\Tr\overline{\left[i\qo+\mu+i\hat\ql-{\cal H}(\vec
	A^{\rm cl}+\vec a_0)+iP \right]\inv}.
\ee
Here $L^2$ is the size of the system and the bar denotes the average
with respect to the action (\ref{defSeff}) with $T\! =\! 1$. Since we know
that the density of a half-filled Landau band is given by 
$B/2\qF_0$ (with $\qF_0$ the flux quantum $h/e$), it
immediately follows from (\ref{CSb}) that near half filling the
Chern-Simons field $b$ must cancel the external field $B$ almost
completely, provided $p\approx 1$. Hence the composite fermion problem
turns into a weak magnetic field problem which can be handled with the
methodology of this paper. This leads to an extension of the actions
(\ref{SsigwithA}) and (\ref{SUwithA}),
\be
	S\naar S_{\rm cs}[a]+S_\qs[A+a]+S_{\rm U}[A+a]
\label{extSkinSU}
\ee
where the $a$ stands for all but the zero-frequency components of the CS
field, and $S_{\rm cs}[a]$ is defined as the 
$\int\! a\!\wedge\! da$ term in
(\ref{2pfluxes}). It is understood that now 
$\qs_{ij}^0\! =\!\qs_{ij}^0(B\! +\! b)$, for which the semiclassical form
(\ref{sigsemiclass}) is a good approximation. Equation (\ref{extSkinSU})
can be written in the form (\ref{SPhiPsiform}) as follows
\bea
\label{CSphipsi}
	S &\naar & S[\Phi+\qf,\qJ+\qj]+S_{\rm cs}[\qf,\qj] \\
	S_{\rm cs}[\qf,\qj] &=& -\qs_{xx}^0\sum_{\qa,n>0}
	\int\!\fr{d^2 q}{(2\pi)^2} 
	n q^2 (\qf^*,\qj^*) M_{\rm cs}
	\left(\matrix{\qf\cr\qj}\right) \\
	M_{\rm cs} &=& \frac{1}{2p\qs_{xx}^0}
	\left(\matrix{0 & -1\cr 1 & 0}\right) \nn
\eea
where we have used the tranverse gauge
$a_i\! =\!\qe_{ij}\prt_j\qj, \; a_\qt\! =\! -\prt_\qt\qf $.

\subsection{Mapping of conductances}
\label{secmap}

The conductances of the composite fermion system can be obtained by
integrating over the CS field $a$. For instance, working with the action
(\ref{responsexy}) with $A\!\naar\! A\! +\! a$, 
one can put $|q|\!\naar\! 0$ first; it
suffices to take the first two terms only. This leads to a mapping of the
composite fermion conductances $\qs_{ij}^0$ to measurable quantities
$\qs_{ij}$, 
\bea
	\qs_{xx}=\frac{\qs_{xx}^0}{(2p\qs_{xx}^0)^2
	+(2p\qs_{xy}^0+1)^2} \hskip0.5cm &;& \hskip0.5cm
	\qs_{xy}=\frac{1}{2p}\left[1-\frac{2p\qs_{xy}^0+1}{(2p\qs_{xx}^0)^2
	+(2p\qs_{xy}^0+1)^2} \right].
\label{mapping}
\eea
This `mapping' is known to be a realization of Sl(2,Z). 
It becomes exact in the limit $\qs_{xx}^0\!\naar\!\infty$, which is the
weak coupling case considered here. Equation (\ref{mapping}) applies
also to the case where $\qs_{xx}^0\!\naar\! 0$ but with a modified
definition of $\qs_{xy}^0$ which is now integrally quantized. 

It is reasonable to assume that (\ref{mapping}) gives a good overall
description provided one is not too close to the critical plateau
transition. In this case one expects the conductances to be broadly
distributed. This then complicates the relation between average and
measured conductances, and (\ref{mapping}) is likely to be affected by
the higher order response terms which gave been neglected in
(\ref{responsexy}) \cite{b5}. 

We can also integrate out the CS field $a_\mu$ working with the full
action (\ref{responsexy}) instead of only the first two terms.
The resulting action for $A_\mu$ has the exact form of
(\ref{responsexy}),
with $\qs_{ij}^0$ replaced by the mapped $\qs_{ij}$ (\ref{mapping}).

\begin{figure}
\begin{center}
\setlength{\unitlength}{1mm}
\begin{picture}(70,70)(0,0)
\put(0,0)
{\epsfxsize=70mm{\epsffile{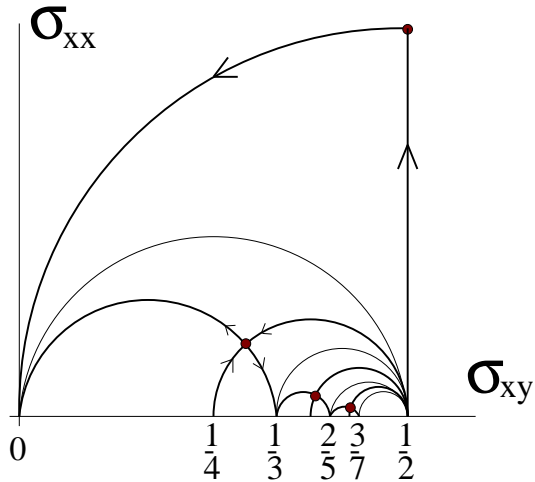}}}
\end{picture}
\caption{Unifying RG diagram for integral and fractional quantum Hall 
states. After Ref~23.}
\label{mappedRG}
\end{center}
\end{figure}

It is important to remark that the Sl(2,Z) mapping (\ref{mapping}) is
neither unique nor universal, but that it depends on  microscopic
details of the system such as disorder. For example, the CS gauge
fields require a different treatment in a theory with slowly varying
potential fluctuations, resulting in a different mapping between
integer and fractional regimes.\cite{fracedge}

\subsection{Internal energy; specific heat}
\label{secinternal}

In order to decide whether the fluctuations in the CS gauge fields are
well-behaved, we next compute the free energy and extract from it the
specific heat. We employ (\ref{CSphipsi}) as well as
(\ref{SPhiPsiform}), (\ref{defM}). For $p\! =\! 1$ we write
\be
	\det[M+M_{\rm cs}]={\cal G}\left[1+(\qo_c\qt)^2+
	\fr{\qo_c\qt}{\qs_{xx}^0}\right]+(2\qs_{xx}^0)^{-2}.
\ee
The contribution to the free energy can be written as
\be
	\qd F=\sum_{\qa,n>0}\int\!\fr{d^2 q}{(2\pi)^2}\ln\left\{
	{\cal G}\left[1+(\qo_c\qt)^2+
	\fr{\qo_c\qt}{\qs_{xx}^0}\right]+(2\qs_{xx}^0)^{-2}
	\right\}.
\ee
In particular, we consider the derivative with respect to temperature
\bea
	\frac{\prt\qd F}{\prt\ln T} &=& 
	\sum_{\qa,n>0}\int\!\fr{d^2 q}{(2\pi)^2}
	\frac{1+(\qo_c\qt)^2+\qo_c\qt/\qs_{xx}^0}
	{{\cal G}[1+(\qo_c\qt)^2+\fr{\qo_c\qt}{\qs_{xx}^0}]
	+(2\qs_{xx}^0)^{-2}}\left[\frac{q^2}{q^2+\qk^2 nU\inv}
	-(\frac{q^2}{q^2+\qk^2 nU\inv})^2\right] \nn\\
	& = &
	\sum_{\qa,n>0}\int\!\fr{d^2 q}{(2\pi)^2}\frac{\qk^2 n
	[1+(\qo_c\qt)^2+\qo_c\qt/\qs_{xx}^0]}
	{\qk^2 n[1+(\qo_c\qt)^2+\fr{\qo_c\qt}{\qs_{xx}^0}]
	+(2\qs_{xx}^0)^{-2}(\qk^2 n+q^2 U)}
	\cdot\frac{q^2}{q^2+\qk^2 n U\inv}.
\eea
This expression is well-behaved in the infrared and for small $\nu_n$
it can be written in the form
\be
	\frac{\prt\qd F}{\prt\ln T}=\sum_{n>0}\nu_n\qr(\nu_n)
\ee
where $\qr(\qo)$ is of order $\qo\ln\qo$ for small $\qo$. However, for free
particles or short-ranged interactions the insertion of CS gauge fields
leads to singular contributions, since by putting 
$U$ constant one finds $\qr(\qo)\!\approx\! |\ln\qo|$ for small
$\qo$. This implies that the CS fields lead to a singular quasiparticle
density of states
\be
	\qr_{qp}(\qe)=\qr(i\qe)+\qr(-i\qe)
\ee
entering the expression for the specific heat \cite{b33}.
The exact meaning of the Chern-Simons gauge field procedure is not
clear in this case. The results nevertheless demonstrate the
fundamental significance of ${\cal F}$-invariance in the problem,
possibly indicating that a new saddlepoint should be found for the
finite range interaction problem.
This, then, shows the importance of the Coulomb interactions.

\ns{Conclusion}
\label{secconclusion}
In this paper we have embarked on the subject of electrodynamic gauge
invariance in the Finkelstein approach to localization and interaction
phenomena. We have found a new symmetry in the problem (${\cal
F}$-invariance) which has fundamental implications in setting up a
unifying theory for the quantum Hall effect. The proposed unifying
theory reconciles Finkelstein's effective action with the topological
concepts of an instanton vacuum and Chern-Simons gauge
theory. Forthcoming analyses will further investigate this theory. 
The second half of this paper has been devoted to the consequences of ${\cal
F}$-invariance for ordinary metals as well as the composite fermion approach
to the half-integer effect.

Future work on this subject will include the tunneling density of
states which will have direct significance for recent experiments.

%goat
\vskip1cm

\noindent
{\bf ACKNOWLEDGEMENTS}\newline
This research was supported in part by
INTAS (grant \#96-0580).

\vskip1cm
\scez\renewcommand{\theequation}{A\arabic{equation}}
\noindent
{\Large\bf Appendix A: Justification of `smallness'}

\vskip0.4cm
\noindent
In order to demonstrate the validity of the `smallness' concept
(section~\ref{sectrunc}, Fig~\ref{figPTQ}) 
let us proceed from the most difficult part of the
action, (\ref{defSeff}),
\be
	\Tr\ln[i\qo+i\hat A_\qt+\hat\mu-\hat{\cal H}+i\hat\ql
	+iT\inv PT]
\ee
and reflect on the possibility of constructing an effective action
$S\eff[A_\mu, T]$ which contains the two distinctly different sets of
field variables $A_\mu$ and $T$ simultaneously. Notice that the
$A_\mu$ are `large' matrices, and the problem is specified to the
question as to why only `small' $T$-rotations are the relevant low
energy excitations. For this purpose we consider the effective actions
$S\eff[A_\mu,T\! =\! 1]$ 
and $S\eff[A_\mu\! =\! 0,T]$ separately. More explicitly,
write
\bea
\label{S[T=1]}
	e^{S\eff[A_\mu,1]} &=& \pathint{[P,\ql]} I[P]\;
	\exp\left\{-\fr{\qb}{2}\int\!\ql\dagg U_0\inv\ql+\Tr\ln[
	i\qo+i\hat A_\qt+\hat\mu-\hat{\cal H}+i\hat\ql+iP]\right\} \\
	e^{S\eff[0,T]} &=& \pathint{[P,\ql]} I[P]\;
	\exp\left\{-\fr{\qb}{2}\int\!\ql\dagg U_0\inv\ql+\Tr\ln[
	i\qo+\hat\mu-\hat{\cal H}_0+i\hat\ql+iT\inv PT]\right\}.
\label{S[A=0]}
\eea
The $S\eff[A_\mu,T\! =\! 1]$ is formally obtained by expanding the Tr ln to
lowest order in powers of the large matrices $A_\mu$ and this means
that complicated infinite sums over frequencies will have to be
performed. The situation for $S\eff[A_\mu\! =\! 0,T]$ is quite different and
one has to follow the procedure of \cite{a4} in order to formally
express this action in terms of the `small' variable $Q$ to lowest
orders in a derivative and temperature expansion.

However, since one is usually interested in the limit of small
momenta, frequencies and temperatures, only a finite number of terms
in $S\eff$ needs to be considered in both cases. The coefficients are
microscopic parameters which are generally given as complex
correlations defined by the underlying theory with plasmon ($\ql$) and
disorder ($P$) variables alone. These coefficients (coupling
constants) of $S\eff[A_\mu,1]$ and $S\eff[0,T]$ are related to one
another by gauge invariance, as will be shown next, and this then
provides the starting point for constructing a complete action
$S\eff[A_\mu,T]$, which is done by `matching' the known results for
both pieces. The details of how to do this are described, in part, in
this paper (Appendix~B) and in a forthcoming paper on the Luttinger
liquid behaviour of disordered edge states in the qHe.

To establish a relation between (\ref{S[T=1]}) and (\ref{S[A=0]}) we
start out by taking a pure gauge for the $A_\mu$ in (\ref{S[T=1]}),
i.e. $A_\mu\! =\!\prt_\mu\qf$, and a `large' matrix field 
$T\! =\! e^{-i\hat\qf}$ in (\ref{S[A=0]}). Writing
\be
	\hat\qf=\sum_{\qa,|n|<M}\qf^\qa_n \Ian
\ee
(\ref{S[T=1]}) and (\ref{S[A=0]}) certainly stand for one and the same
thing. Next, we make this statement useful by showing that the large
rotation $T\! =\! e^{-i\hat\qf}$ can be replaced by an equivalent rotation
($t$) which is small. For this purpose we write
\be
	\hat\qf=\hat\qf_t+\hat\qf_l
\ee
where $\hat\qf_l$ is block diagonal (nonzero only in the $++$ and $--$
Matsubara blocks) and $\hat\qf_t$ is block off-diagonal (nonzero in
the $+-$ and $-+$ blocks). Now write
\be
	T=e^{-i\qk\hat\qf}=e^{-i\qk\hat\qf_l}t(\qk)
\label{Tsplit}
\ee
where the parameter $\qk$ formally serves as an expansion
parameter. The $t(\qk)$ can be written as a series in powers of $\qk$
as follows:
\be
	t(\qk)=\exp\sum_{n=1}^\infty \fr{i\qk^n}{n!}\hat x_n
\ee
where
\bea
	\hat x_1=\hat\qf_t \hskip0.3cm &
	\hat x_2=[\hat\qf_t,\hat\qf_l] \hskip0.3cm &
	\hat x_3=[[\hat\qf_t,\hat\qf_l],\hat\qf_l]\;\;\;\; \ldots
\eea
The important point is that the $\hat x_n$ are all block off-diagonal
matrices and their `size' in frequency space increases linearly in
$n$. 
It serves our purpose to truncate the series
beyond small orders in $\qk$ such that $t(\qk)$ satisfies the
condition of `smallness'. The statement (\ref{Tsplit}) now effectively
turns into a separation of large components $e^{-i\qk\hat\qf_l}$ and
small components $t(\qk)$. The large components can be absorbed into a
redefinition of the $P$-field which leads to the statement
\be
	T\inv PT=t\inv(\qk)Pt(\qk)
\ee
or, equivalently, 
\be
	S\eff[\qk\prt_\mu\qf,1]=S\eff[0,t(\qk)].
\label{gaugeAT}
\ee
This procedure can be extended as follows: Suppose we have found
$S\eff[A_\mu,T]$ from a `matching' procedure as mentioned above. A
useful check upon this result is obtained by a generalization of
(\ref{gaugeAT}),
\be
	S\eff[A_\mu+\qk\prt_\mu\qf,t(\qk)]=S\eff[A_\mu,1].
\label{gengaugeAT}
\ee

\vskip1.5cm

\scez\renewcommand{\theequation}{B\arabic{equation}}
\noindent
{\Large\bf Appendix B}

\vskip0.4cm
\noindent
In order to give an example of the matching procedure (Appendix A), we
derive an effective action in the plasmon field $\ql$ and the matrix
field variable $T$. Define $S\eff[\ql,T\!=\!1]$ and
$S\eff[\ql\!=\!0,T]$ as follows,
\bea
\label{lambdacase}
	e^{S\eff[\ql,1]} &=& \pathint{P}I[P]
	\exp\{-\fr{\qb}{2}\int\! \ql\dagg U_0\inv \ql
	+\Tr\ln[i\qo+\mu-\hat{\cal H}_0+i\hat\ql+iP]\} \\
\label{freecase}
	e^{S\eff[0,T]} &=& \pathint{P}I[P]
	\exp\{\Tr\ln[i\qo+\mu-\hat{\cal H}_0+iT\inv PT]\}.
\eea
The idea is to construct $S\eff[\ql,T]$ from a detailed knowledge of
Eqs. (\ref{lambdacase}) and (\ref{freecase}). Notice that
(\ref{freecase}) is precisely the free particle problem. Eq
(\ref{freecase}) is evaluated by writing the Tr ln[ ] as
\be
	\Tr\ln[iT\qo T\inv+\mu-T\hat{\cal H}_0 T\inv +iP]=
	\Tr\ln[\mu-\hat{\cal H}_0+iP + X],
\ee
where $X\!=\!iT\qo T\inv\!-\!T[\hat{\cal H}_0,T\inv]$ is a small
parameter. An expansion in powers of $X$ leads to a systematic
expansion of $S\eff[\ql\!=\!0,T]$ in powers of the gradient and
temperature. The result can be expressed in the field variable $Q$ as follows
\be
	S\eff[0,T]=S_\qs[Q]+\fr{2\pi}{\qb}\pi\qr_0\Tr\qh Q +\cdots
\label{Sefffree}
\ee
where $\qr_0$ is the free particle density of states, 
equal to $dn/d\mu$ in this case, which can be written as
\bea
	\qr_0=-\fr{1}{2\pi i}\left\langle G^{\qa\qa}_{n_1 n_1}(x,x)
	-G^{\qa\qa}_{n_2 n_2}(x,x) \right\rangle_{\rm av} 
	\hskip0.5cm &;& \hskip0.5cm
	G(x,x')=\langle x|(\mu-\hat{\cal H}_0+iP)\inv |x'\rangle
\label{defqr0}
\eea
where the average is with respect to the theory of (\ref{freecase})
with $T\!=\!1$ and $\qo\!=\!0$. In (\ref{defqr0})
the indices are kept fixed with $n_1\! >\! 0$ and $n_2\! <\! 0$ as usual.
Eq. (\ref{defqr0}) is identical to the more familiar expression for
$\qr_0$, as can be seen from the standard rules of replica field theory.
More specifically, for quantities like (\ref{defqr0}) which involve
unmixed averages over the positive and negative blocks of $P$, one can
transform the problem back and trade in the $P$-integral for the
average over the original random potential $V(\vec x)$.
On the other hand, we write (\ref{lambdacase}) as an expression in
powers of $\ql$. The result to lowest order in $\ql$ can be written as
\be
	S\eff[\ql,1]=-\fr{\qb}{2}\sum_{n\qa}\int\! \ql^\qa_{-n}
	 U_0\inv \ql^\qa_n
	-\half\sum_{\qa\qb}\sum_{n,m}\int \!
	\ql^\qa_n M^{\qa\qb}_{nm}\ql^\qb_m
\label{qlexpansion}
\ee
where
\bea
	M^{\qa\qb}_{nm}(x,x') & =& -\tr\left\langle\hat G(x,x')\;
	{\rm I}^\qb_m\hat G(x',x)\;\Ian\right\rangle_{\rm av} 
	+\left\langle\tr[\hat G(x,x)\;{\rm I}^\qb_m]\tr[\hat
	G(x',x')\;\Ian]\right\rangle_{\rm cum}\nn
	\\
	\hat G(x,x') &=& \langle x|(i\qo+\mu-\hat{\cal H}_0+iP)\inv
	|x'\rangle.
\label{B6}
\eea
The subscript `cum' stands for the cumulant average with respect to
(\ref{lambdacase}) with $\ql\!=\!0$.
Notice the subtle difference in the expansions of (\ref{lambdacase})
and (\ref{freecase}), in that the $\qo$-matrix is treated differently
in (\ref{B6}) and (\ref{defqr0}), leading to different propagators
$\hat G$ and $G$, respectively. 

The matrix elements $M^{\qa\qb}_{nm}$ in (\ref{qlexpansion}) 
can be simplified by making use of the fact that the expectations of
$\hat G$ (\ref{B6}) are invariant under unitary
transformations. Specifically, (\ref{B6}) is invariant under the
replacement 
\be
	\hat G^{\qa\qb}_{nm}(x,x')=[U\inv \hat G(x,x') U]^{\qa\qb}_{nm},
\label{B7}
\ee
where $U$ is diagonal in the Matsubara frequency index,
$U^{\qa\qb}_{nm}=\qd_{nm}U_m^{\qa\qb}$.
It is then straightforward to show that $M^{\qa\qb}_{nm}$ must be of
the general form
\be
	M^{\qa\qb}_{nm}(x,x')=\qd^{\qa\qb}\qd_{n+m,0}M_1(x-x',\qo_n)
	+\qd_{m,0}\qd_{n,0}M_0(x-x').
\label{Mxx'}
\ee
The two different terms in (\ref{Mxx'}) have an entirely different
meaning and they are going to be treated quite differently in what
follows. First, the quantity $M_1$ can be expanded in a series
expansion in small momenta (gradients) and frequencies. To lowest
order we have
\be
	M_1(x-x',\qo_n)=\qb\qr_0\qd(x-x')+\cdots.
\label{B9}
\ee
Here, $\qr_0$ can be identified as the exact free particle density of
states (\ref{defqr0}). The $\cdots$ in (\ref{B9}) stands for all the
higher order terms in frequency and derivatives. They become important
only when higher dimensional operators in $Q$ (represented by $\cdots$
in (\ref{Sefffree})) are taken into account. 

Next, the zero frequency quantity $M_0(x-x')$ in (\ref{Mxx'}) can be
identified as the `mean field' result for the density fluctuation
correlation 
$\overline{\qd n(x) \qd n(x')}$, where 
$\qd n(x)\!=\! n(x)\!-\! \overline{n(x)}$ and the bar denotes the
ensemble average. We shall see that full $M_0(x-x')$ (i.e. without
momentum expansion) completely decouples from the effective action
procedure and is, in fact, immaterial. 

The idea then is to find the `match' between the different series
expansions (\ref{qlexpansion}--\ref{B9}) and (\ref{Sefffree}).
Schematically, this is given by
\bea
	S\eff[\ql,T] &=& -\fr{\qb}{2}\sum_{n\qa}\int\! \ql^\qa_{-n} 
	U_0\inv \ql^\qa_n-\half\sum_{\qa\qb}\int\ql_0^\qa M_0 \ql_0^\qb
	+S_\qs[Q]
	\nn\\ &&
	+\fr{2\pi}{\qb}\pi\qr_0
	\int\!\left[\tr\qh Q+\fr{\qb}{2\pi}\sum_{n\qa}\ql^\qa_n
	\tr\Ian Q
	-(\fr{\qb}{2\pi})^2\sum_{n\qa}\ql^\qa_{-n}\ql^\qa_n\right]. 
\label{match}
\eea
It can be shown that (\ref{match}) satisfies (\ref{gengaugeAT}).
Eq. (\ref{match}) therefore is the desired result. Moreover, comparison
with (\ref{transetaQ}) shows that (\ref{match}) is ${\cal F}$-invariant.
Next, by making the appropriate shift 
\[
	\ql^\qa_n\naar
	\ql^\qa_n+\fr{\pi}{\beta}\rho_0(U_0^{-1}+\rho_0)^{-1}\tr\Iamn Q,
\]
the final result
decouples such that we have
\be
	S\eff(\ql,T)=S\eff[\ql]+S_\qs[Q]+S_{\rm F}[Q]+S_{\rm U}[Q]
\ee
where
\bea
	S\eff[\ql]&=& -\fr{\qb}{2}\int\ql\dagg(U_0\inv +\qr_0)\ql -\half
	\sum_{\qa\qb} \int\ql_0^\qa M_0 \ql_0^\qb \nn\\
	S_{\rm F}[Q]&=& \fr{\pi^2}{2\qb}\qr_0 \int\left[ \sum_{n\qa}\tr\Ian Q\;
	\tr\Iamn Q+4\tr\qh Q\right] \nn\\
	S_{\rm U}[Q]&=&-\fr{\pi}{\qb}\sum_{n\qa}\int \tr(\Ian Q) U\inv
	\; \tr(\Iamn Q). 
\label{decoupled}
\eea
This is precisely the form written in (\ref{SeffQ}--\ref{SU}).
The procedure that has taken us from (\ref{lambdacase},\ref{freecase})
to (\ref{decoupled}) can be systematically extended to include higher
orders. This means that terms of higher dimension in (\ref{Sefffree})
and (\ref{B9}) as well as higher powers of $\ql$ in
(\ref{qlexpansion}) can be taken into account. The extended procedure
leads to renormalization (in the Fermi liquid sense) of the parameters
in (\ref{decoupled}) and it generates higher dimensional operators in
$Q$ as well. A detailed analysis will be reported elsewhere.\cite{paperV}

%We mention that the procedure can be extended to higher
%orders\cite{paperV}, which 
%means that higher orders in $\ql$ can be considered in
%(\ref{qlexpansion}) along with higher order operators in $Q$
%(proportional to $\qb^{-2}$) in (\ref{Sefffree}). These higher order
%contributions renormalize (in the Fermi liquid manner)
%the coefficients in (\ref{decoupled})
%and they generate higher order operators as well.


\begin{thebibliography}{99}


\bibitem[*]{Misha} Permanent address: Russian Research Center 
"Kurchatov Institute", 
Kurchatov sq. 1, 123182 Moscow, Russia

\vskip2mm

\bibitem[\dagger]{foot1} 
The invariance of $S_{\rm F}$ can be easily understood
by first writing 
\be
	S_{\rm F}[WQW\inv]=
	\half z\fr{\pi}{\qb}{\sum_{\qa n}}'\Tr[W\inv \Ian W,Q]
	[W\inv \Iamn W,Q]
\ee
and then splitting 
\be
	W\inv \Ian W=\Ian+W\inv [\Ian,W].
\label{deltaI}
\ee
The second term on the r.h.s. has nonzero matrix elements only in the upper
left-hand and lower right-hand corners, i.e. in a `small' neighborhood of 
the extreme diagonal components 
$(-N_{\rm max}',-N_{\rm max}')$ and
$(N_{\rm max}' \! -\! 1,N_{\rm max}' \! -\! 1)$, 
as can be seen by expanding the exponential
form of $W$ in powers of the I-matrices. Therefore, the
second term in (\ref{deltaI}) commutes with $Q$ and we have
$S_{\rm F}[WQW\inv]\!=\!S_{\rm F}[Q]$.

\vskip2mm

%1
\bibitem{c3} For a review, see {\it The Quantum Hall Effect}, edited
by R.E. Prange and S.M. Girvin (Springer-Verlag, Berlin, 1987)

%2
\bibitem{a2} H. Levine, S. Libby and A.M.M. Pruisken, {\it
Phys. Rev. Lett.} {\bf 51}, 20 (1983)

%3
\bibitem{a4} A.M.M. Pruisken, {\it Nucl. Phys.} {\bf B235} [FS11], 277
(1984)

%4
\bibitem{a8} See, e.g., A.M.M. Pruisken and H.P. Wei, {\it AIP
Conf. Proc.} {\bf 286}, 159 (1994)

%5
\bibitem{a6} A.M.M. Pruisken, {\it Phys. Rev.} {\bf B32}, 2636 (1985)

%6
\bibitem{a9} A.M.M. Pruisken, {\it Phys. Rev. Lett.} {\bf 61}, 1297 (1988)

%7
\bibitem{a11} H.P. Wei, D.C. Tsui, M.A. Palaanen and A.M.M. Pruisken,
{\it Phys. Rev. Lett.} {\bf 61}, 1294 (1988)

%8

\bibitem{c5} R. Laughlin, {\it Phys. Rev. Lett.} {\bf 50}, 1395 (1983)


\bibitem{b18} R.B. Laughlin, M.L. Cohen, J.M. Kosterlitz, H. Levine,
S.B. Libby and A.M.M. Pruisken, {\it Phys. Rev.} {\bf B32}, 1311 (1985)


\bibitem{PruiskenBaranov} A.M.M. Pruisken and M.A. Baranov,
{\it Europhys. Lett.} {\bf 31}(9), 543 (1995)


\bibitem{b8} A.M. Finkelstein, {\it JETP Lett.} {\bf 37}, 517 (1983);
{\it Soviet Phys. JETP} {\bf 59}, 212 (1984);
{\it Physica} {\bf B197}, 636 (1994)







\bibitem{Wilczek82} F. Wilczek, {\it Phys. Rev. Lett.} {\bf 48}, 1144 (1982)


\bibitem{c11} A.L. Fetter, C.B. Hanna and R.L. Laughlin, {\it
Phys. Rev.} {\bf B39}, 9679 (1989); {\bf 43}, 309 (1991)


\bibitem{c12} A. Lopez and E. Fradkin, {\it Phys. Rev.} {\bf
B44}, 5246 (1991); {\it Phys. Rev. Lett.} {\bf 69}, 2126 (1992) 


\bibitem{c13} S.C. Zhang. H. Hansson and S. Kivelson, {\it
Phys. Rev. Lett.} {\bf 62}, 82 (1989); See S.C. Zhang, {\it
Int. J. Mod. Phys.} {\bf B6}, 25 (1992)


\bibitem{c14} S. Kivelson, S.C. Zhang and D.H. Lee, {\it Phys. Rev.}
{\bf B46}, 2223 (1992)


\bibitem{b5} M.H. Cohen and A.M.M. Pruisken, {\it AIP Conf. Ser.} {\bf
286}, 205 (1993); {\it Phys. Rev.} {\bf B49}, 4593 (1994)


\bibitem{c15} B.I. Halperin, P.A. Lee and N. Read, {\it Phys. Rev.}
{\bf B47}, 7312 (1993)



\bibitem{c10} J.K. Jain, {\it Phys. Rev. Lett.} {\bf 63}, 199 (1989)


\bibitem{b30} This was one of the main results of Lopez and Fradkin
\cite{c12}, who discarded the edge states.


\bibitem{b27} X.G. Wen, {\it Phys. Rev.} {\bf B41}, 12838 (1990)


\bibitem{b31} A.M.M. Pruisken, B. \v{S}kori\'{c} and M.A. Baranov,
``Handling gauge invariance in the theory of the quantum Hall effect
III: The instanton vacuum and chiral edge physics'', ITFA preprint
98-17, cond-mat/9807241, to appear in {\it Phys. Rev.} B



\bibitem{PrSch} A.M.M. Pruisken and K. Schoutens, {\it Phil. Mag.}
{\bf B76} 807 (1997)

\bibitem{fracedge} B. \v{S}kori\'{c} and A.M.M. Pruisken,
``The fractional quantum Hall effect: Chern-Simons mapping, duality,
Luttinger liquids and the instanton vacuum'', cond-mat/9812437,
accepted for publication in {\it Nucl. Phys.} B


\bibitem{b32} A similar phenomenon is well known to occur in ordinary
Fermi liquid theory where large frequency sums or non-Fermi level
quantities contribute to the effective Landau parameters.


\bibitem{b33} M.A. Baranov, A.M.M. Pruisken and B. \v{S}kori\'{c},
''Handling gauge invariance in the theory of the quantum Hall effect
II: Perturbative results'', ITFA preprint 97-48, cond-mat/9712323,
to appear in {\it Phys. Rev.} B

\bibitem{paperV} M.A. Baranov and A.M.M. Pruisken,
to be published. A simplified version of this theory is given in
Appendix~B.

\bibitem{Nozieres} D. Pines and P. Nozi\`{e}res, {\it The Theory of
Quantum Liquids: Volume 1} (W.A. Benjamin, Inc., New York 1966)

\end{thebibliography}
\end{document}